\documentclass{aa}  
\usepackage{graphicx}
\usepackage{placeins}
\usepackage{xcolor}
\usepackage{mathtools}
\usepackage{bbm}
\usepackage{placeins}

\usepackage{url}
\usepackage{amsmath}
\usepackage{multirow}
\usepackage[colorlinks = true,
           linkcolor = blue,
           urlcolor  = blue,
           citecolor = blue,
           anchorcolor = blue]{hyperref}
\usepackage[labelfont={bf,color=blue}]{caption}
\usepackage{mhchem}
\usepackage{txfonts}
\newcommand{\orcit}[1]{\protect\href{https://orcid.org/#1}{\protect\includegraphics[width=8pt]{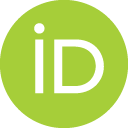}}}

\newcommand{\Ha}{\ensuremath{\rm H\alpha}\xspace}

\newcommand{\Gaia}{\textit{Gaia}\xspace}

\newcommand{\userlabel}{\texttt{user\_label}\xspace}
\newcommand{\msl}{\texttt{manual\_spectral\_label}\xspace}

\newcommand{\referee}[1]{#1}

\begin{document}

   \title{{\tt XP-TEAL}: \textit{Gaia} XP Tool for Emission and Absorption Lines \\
          II. \textit{Gaia}-HELIX catalogue of \ensuremath{\rm H\alpha} emitters in \textit{Gaia} BP/RP spectra}
   
   \author{
   S. Malhotra\orcit{0000-0002-5509-0168}\inst{1,2,3} \and  
   M. Weiler\orcit{0000-0002-3007-3927}\inst{1,2,3} \and
   F. Anders\orcit{0000-0003-4524-9363}\inst{1,2,3} \and
   J. M. Carrasco\orcit{0000-0002-3029-5853}\inst{3,2,1} \and
   G. Garcia-Moreno\orcit{0009-0001-4098-7706}\inst{2,3}\and  
   C. Jordi\orcit{0000-0001-5495-9602}\inst{1,4},\\ 
   N. Blagorodnova\orcit{0000-0003-0901-1606}\inst{1,2,3}\and 
   A. Castro-Ginard\orcit{0000-0002-9419-3725}\inst{1,2,3}\and
   J. Casta\~{n}eda\orcit{0000-0001-7820-946X}\inst{5,2,1}\and 
   F. Boix\inst{1,2,3}\and 
   X. Luri\orcit{0000-0001-5428-9397}\inst{1,2,3}}

\authorrunning{Sagar Malhotra et al.}
\titlerunning{\Ha emitters in \Gaia XP spectra}

   \institute{
   Institut d'Estudis Espacials de Catalunya (IEEC), Edifici RDIT, Campus UPC, 08860 Castelldefels (Barcelona), Spain
\and   
   Institut de Ci\`encies del Cosmos (ICCUB), Universitat de Barcelona (UB), Mart\'{\i} i Franqu\`es 1, E-08028 Barcelona, Spain
   \and
   Departament de F\'{\i}sica Qu\`antica i Astrof\'{\i}sica (FQA), Universitat de Barcelona (UB), Mart\'{\i} i Franqu\`es 1, E-08028 Barcelona, Spain
   \and
   Institut d'Estudis Catalans, c. del Carme, 47, E-08001 Barcelona, Spain
   \and
   DAPCOM Data Services, c. dels Vilabella, 5-7, 80500 Vic, Barcelona, Spain
\\   
\email{\{sagar,mweiler\}@fqa.ub.edu}
             }

   \date{Received 17 June 2026; Accepted 10 August 2026}

  \abstract
   {With \textit{Gaia}'s third Data Release (DR3), low-resolution BP/RP (XP) spectra became available for more than 219 million sources. In previous work, we developed a fast method for detecting absorption and emission lines while measuring their equivalent widths directly from the XP spectral coefficients.} 
   {We conduct a systematic search for \ensuremath{\rm H\alpha} emitters in the \textit{Gaia} DR3 XP spectra.}
   {We measured the equivalent width of the Balmer \ensuremath{\rm H\alpha} line for all sources with \textit{Gaia} XP spectra and selected an initial sample of 556\,100 sources depicting an \ensuremath{\rm H\alpha} emission-like feature at $2\sigma$ level or above. Using a semi-supervised classifier complemented by manual spectral labelling, we removed sources that merely masqueraded as emission-line stars.}
   {We find that 95\% of the initial candidates are most likely cool M-type stars, where a molecular absorption band gives rise to a local maximum that can mimic \ensuremath{\rm H\alpha} emission. Subsequently, 28\,394 sources were flagged as bona fide \Ha emitters using the \textit{Gaia} DR3 XP spectra. The majority are Be-like candidates (26\,287), M-type \referee{emission-line} stars (717), Herbig~Ae/Be candidates (525), quasi-stellar objects (204), Wolf-Rayet stars (177), cataclysmic variables (51), and carbon stars (18), among other sources. Our equivalent widths and stellar classifications agree well with results from higher-resolution studies and recent \textit{Gaia} XP spectra-based catalogues. Ground-based spectroscopic follow-up for some of the newly found \ensuremath{\rm H\alpha} emitters reinforces our confidence in the catalogue.}
  {}


   \maketitle
%

\section{Introduction \label{sec:introduction}}

Stars of various spectral types can display hydrogen emission lines in their spectra \citep[e.g.][]{Merrill1950, Wackerling1970, Kogure2007, Yu2025}. The presence, strength, profile, and variability of these lines give valuable indications about the physical mechanisms driving their formation, and ultimately, about the stellar evolution in different spectral classes \citep[e.g.][]{Underhill1990, Kovtyukh2011, Labadie-Bartz2017, Garcia-Moreno2026}.

Numerous studies have searched for emission-line stars using narrow-band imaging and large-scale spectroscopic surveys \citep{Henize1976, Kohoutek1999, Drew2005, Hou2016, Vioque2018, Cotar2021}. Recently, \citet{Vioque2020} combined \Gaia Data Release 2 (DR2; \citealt{GaiaCollaboration2018}) astrometry and multi-band photometry in a machine-learning pipeline, delivering thousands of new pre-main-sequence and classical Be candidates. The release of the \Gaia DR3 low-resolution BP/RP (XP) spectra for $\sim$219 million sources \citep{Vallenari2023} transformed this picture: the {\it Apsis} pipeline \citep{Creevey2022} introduced the Extended Stellar Parametrizer for Emission-Line Stars \citep[ESP-ELS;][]{Fouesneau2023}, which flagged 57\,511 sources in seven ELS classes (Be, Herbig~Ae/Be, T~Tauri, dMe, WC, WN, and PNe), and the Extended Stellar Parametrizer for Hot Stars (ESP-HS) for $\sim$2.3 million stars hotter than 7\,500~K. These classifications were independently validated by \citet{Shridharan2022} and have since enabled numerous targeted studies of specific stellar populations (e.g. \citealt{Ball2025,  An2025}; \referee{\citealt{Mulato2025, Mulato2026}}; \citealt{Lailey2026,Carrasco2026}).

We analyse the full set of sources with \Gaia DR3 low-resolution XP spectra to search for and characterise \Ha emitting stars. In Sect.~\ref{sec:method}, we briefly describe the \Gaia XP spectra and explain how we use the {\tt XP-TEAL} tool \citep{weiler2023, Carrasco2026} to find 556\,100 candidate \Ha emitters in a blind search over the full dataset. Further, we develop a semi-supervised classifier that helped us remove objects that only masquerade as emission-line stars and classify the reliable objects into spectral sub-classes. In Sect.~\ref{sec:cat} we provide an overview of the catalogue and various stellar populations. In Sect.~\ref{sec:literature} we present an extensive comparison with previous literature on \Ha emitters in terms of completeness and contamination of our classifications in addition to the reliability of our estimated equivalent widths. In Sect.~\ref{sec:followup} we show that ground-based follow-up spectroscopic observations for some of the brightest newly found \Ha emitters reinforce our confidence in the catalogue. Finally, Sect.~\ref{sec:conclusions} summarises our results and conclusions.

\section{Method \label{sec:method}}

The third data release (DR3; \citealt{Vallenari2023}) of the \Gaia mission \citep{Prusti2016} delivered low-resolution spectra for approximately 219 million astronomical sources \citep{DeAngeli2023}, most of them with $G$-band apparent magnitudes brighter than $17.65$ mag. The blue photometer (BP) covers wavelengths between approximately 330 nm and 680 nm, and the red photometer (RP) spans 640 nm to 1050 nm \citep{Carrasco2021}. The spectral resolution of the two instruments is below $R\sim100$ and varies slightly with wavelength. For simplicity, we use "XP" to refer collectively to \Gaia BP and RP spectra.

In \Gaia DR3, XP spectra are represented as coefficients for a series of basis functions \citep{Carrasco2021, DeAngeli2023} that can be used to construct internally and externally calibrated spectra. Internally calibrated spectra express flux in terms of photoelectrons per second per pseudo-wavelength unit (the pseudo-wavelength is essentially a proxy for the CCD pixel scale) within the \Gaia aperture. Externally calibrated spectra, combining BP and RP data, provide flux in physical units (energy per unit of time, wavelength, and detector area) after deconvolving the internal basis functions of the low-resolution spectra with the line spread function (LSF; \citealt{Montegriffo2023}).

For internal spectra, the basis consists of shifted and scaled Hermite functions, optimised in \Gaia DR3 through orthogonal transformations to better match typical spectral shapes \citep{Carrasco2021}. Hermite functions offer several computational advantages: derivatives of their linear combinations remain linear combinations, enabling efficient computation via matrix operations. Additionally, the roots of such combinations correspond to eigenvalues of a companion matrix, allowing us to rapidly identify extrema in internally calibrated XP spectra. Hence, by using the internally calibrated spectra, we also avoid the deconvolution step, which is prone to produce spurious features in the externally calibrated spectra \citep{weiler2023, Montegriffo2023, Huang2024}. For these reasons, we used internally calibrated spectra throughout this paper.

The properties explained above were leveraged by \citet{weiler2023} to develop the XP Tool for Emission and Absorption Lines ({\tt XP-TEAL}). The main idea is that spectral lines appear as local extrema in the spectra or their second derivatives, depending on line strength and continuum gradients. By modelling the LSF as a linear combination of Hermite functions and enforcing smoothness (vanishing higher-order derivatives at line centres), equivalent widths and their uncertainties can be derived efficiently. The method was successfully tested on open cluster member stars in our recent companion paper \citep{Carrasco2026}.

\subsection{{\tt XP-TEAL} run on all \Gaia DR3 XP spectra}\label{subsec:xp_teal}

\begin{figure}
    \centering
\includegraphics[width=0.99\columnwidth]{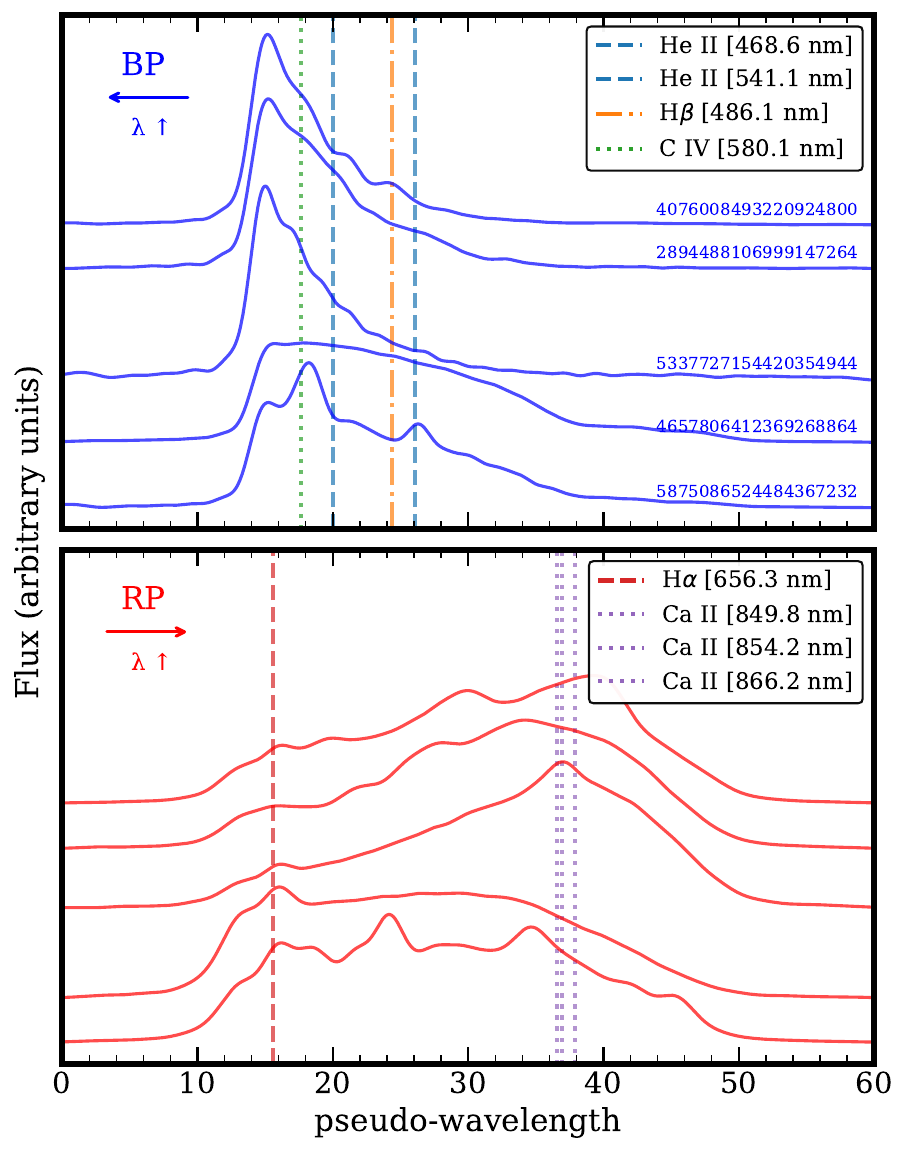}
    \caption{\Gaia DR3 internal spectra (pseudo-wavelength on the x-axis) in the BP (\textit{top}) and RP (\textit{bottom}) bands for 5 stars from the initial sample of 556\,100 sources. The vertical dashed lines mark the expected positions of some spectral lines, as indicated in the legend, and the spectra in the top panel are labelled with their corresponding \Gaia DR3 source identifiers. Arbitrary vertical offsets have been applied for clarity. The arrows indicate the sense of increasing wavelength in each filter.}
    \label{fig:internal_spectra_example}
\end{figure}

With the aim to compile a catalogue of \Ha emitters in \Gaia DR3, we applied {\tt XP-TEAL} to the full \Gaia DR3 XP dataset with more than 219 million stars \citep{DeAngeli2023}, requiring a total of $\sim$11\,400 CPU hours on the MareNostrum~5 supercomputer at the Barcelona Supercomputing Center. All computations were carried out under the assumption of narrow spectral lines (i.e. the intrinsic width of a spectral line is small compared to the width of the LSF) and considering the sources with the \Ha line-width parameter, $\omega$ (measure of the distance between the inflection points associated with a local extremum; see Sect.~9 of \citealt{weiler2023}), between 0.8 and 1.1. Furthermore, we applied a constraint on the signal-to-noise ratio of the \Ha line (S/N (\Ha)) for the estimated equivalent width, EW, such that $\frac{EW(\text{H}\alpha)}{\sigma_{EW(\text{H}\alpha)}} \geq 2$, where $\sigma_{EW(\text{H}\alpha)}$ is the uncertainty on $EW(\text{H}\alpha)$, to avoid very weak or spurious detections. This ensured that we only included sources with \Ha emission feature\footnote{We note the unconventional sign usage for $EW(\text{H}\alpha)$; we use positive for emission and negative for absorption.} in their \Gaia XP spectra, resulting in a sample of 556\,100 sources. Of these, 174\,084 sources have S/N (\Ha)~$\geq3$ while 382\,016 have S/N (\Ha) $\in[2, 3)$. All except one source (\texttt{Gaia DR3 4042380857826322176}) have at least two extrema in their observed spectrum or its second derivative, that is, at least one in addition to that associated with the \Ha line. We expect this S/N~(\Ha) to depend on the \Gaia scanning law (number of observations) and to vary for different stellar populations, as shown in Fig.~\ref{fig:scanning_law_snr}. An example of the \Gaia~DR3 internal spectra for 5 sources that are included in our initial sample of 556\,100 sources is shown in Fig.~\ref{fig:internal_spectra_example}. For the RP band, the wavelength increases with increasing pseudo-wavelength, and vice versa for the BP band.

\newpage
\subsection{Classification of candidate \Ha emitters}\label{subsec:nn_classification}

\begin{figure*}
    \centering
\includegraphics[width=0.99\textwidth]{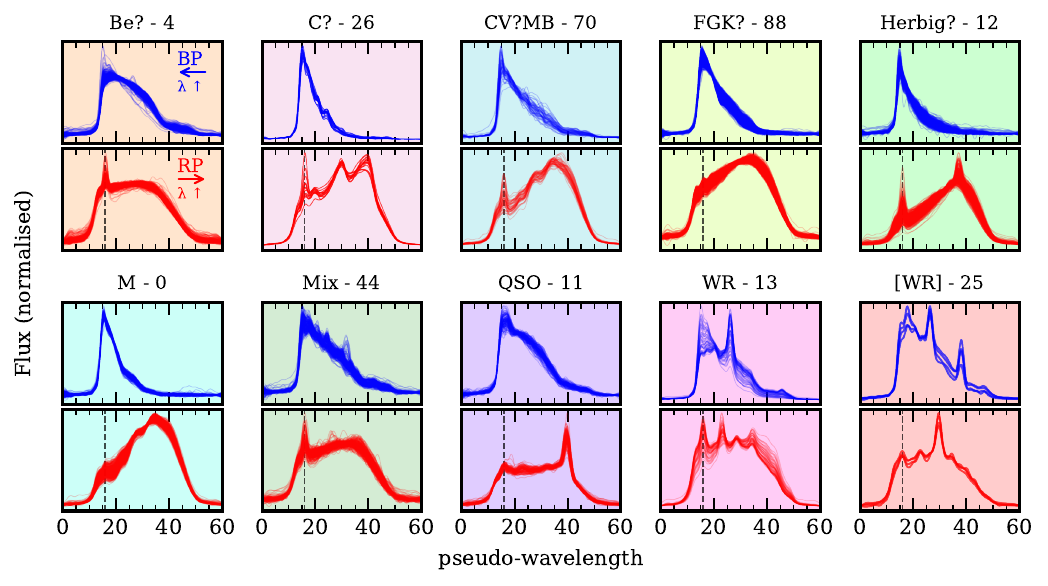}
    \caption{Example \userlabel classes of our classification of \Ha emitters for each \msl. A maximum of 1000 spectra per class is shown to illustrate the similarity and variance of the spectra within one class. The title of each subplot indicates \msl\xspace - \userlabel and the \Ha position is shown by the dashed black line. A complete version of this figure depicting each \userlabel and \msl is shown in Fig.~\ref{fig:NN_user_label_subplot}.}
    \label{fig:specific_user_labels}
\end{figure*}

Figure~\ref{fig:internal_spectra_example} demonstrates the variety in spectral shapes of the sources for which {\tt XP-TEAL} returned an emission in \Ha. While many of these sources are genuine \Ha emitters, this sample is heavily contaminated by stars with a local maximum that only appears similar to an \Ha emission line in the low-resolution RP spectra. This is particularly the case for M-type stars, where molecular features such as TiO absorption bands produce a broad local minimum. In combination with the generally rising RP continuum towards longer pseudo-wavelengths, these minima can produce statistically significant local maxima on their short-wavelength side, frequently near the \Ha position. To reject these false positives and classify genuine \Ha emitters into physically meaningful subclasses, we carried out a semi-supervised classification of all candidate emitter spectra.

We first used a sample of stars with a higher S/N (\Ha) ($\geq 3$) to construct a training dataset for the subsequent classification using a neural network (NN). This dataset was built largely through manual labelling, where stars with similar XP spectra were assigned a common \userlabel. New classes were defined based solely on the morphology of the observed XP spectra without incorporating underlying physical considerations (e.g. differences arising from reddening among stars of the same spectral type), with the aim of enabling the NN to better capture subtle variations in the XP spectra. This approach was adopted to enhance the NN classification performance, reduce misclassification, and improve the identification of outliers. The final training sample consisted of 89 distinct classes (i.e. \userlabel is an integer between 0 and 88), with the number of sources\footnote{About 15 sources are caused by poor calibration or blending from another nearby source, but we kept these sources in our sample for completeness.} in each class ranging from 1 to 66\,284. We trained the artificial NN to classify the XP spectra using the first 30 truncated and optimised \citep{Carrasco2021} XP coefficients, the observed $G_{\rm BP}-G_{\rm RP}$ colour ($G_{\rm BP}$ and $G_{\rm RP}$ are the observed mean magnitudes in the \Gaia BP and RP bands, \citealt{Riello2021}) together with some features describing the shape of spectra and their derivatives. For this last set of features, we computed the number of roots and the integral of the absolute value of the ${n}^{th}$ \referee{[n $\in$ (1, 2, 3)]} derivative of spectra in a particular pseudo-wavelength range. These features helped the NN to distinguish between a cool star with molecular bands and a hot star (e.g. B-type) with emission lines (see Fig.~\ref{fig:shap}). A detailed description of the various features used to train the NN, along with their relative importance in model prediction, is provided in App.~\ref{app:NN_features}. The NN consisted of three hidden layers (256, 128, and 64 neurons) with rectified linear unit (ReLU) activation functions, batch normalisation, dropout regularisation, and L2 weight decay. After it was trained, the NN was applied to the remaining 382\,016 sources with S/N (\Ha)~$\in$~[2, 3), and the final \userlabel for each source was assigned by selecting the class with the highest predicted probability from the network. We manually flagged obvious misclassifications in each \userlabel while assigning the corresponding sources a \texttt{classification\_flag} of $0$. Finally, since our \userlabel is not immediately translatable to an astrophysical class, we combined several {\tt user\_labels} that are most likely to correspond to a similar object type and assigned them a human-readable class, called \msl (see Tab.~\ref{tab:helix_summary}). This was also aided by using the most referred to object type in SIMBAD \citep{Wenger2000} for each \userlabel. 

We converged on a total of $11$ \msl classes: M-type ({\tt M}), Be-like ({\tt Be?}), Herbig~Ae/Be-like ({\tt Herbig?}), quasi-stellar objects ({\tt QSO}), Wolf-Rayet ({\tt WR}), central stars of planetary nebulae depicting WR-like spectral features ({\tt [WR]}), sources with flux excess from a nearby source ({\tt Blending}), planetary nebulae ({\tt PN}), likely carbon stars ({\tt C?}), likely FGK stars ({\tt FGK?}), and cataclysmic variables depicting molecular bands ({\tt CV?MB}). We also added two other classes: a class composed of a mix of sources (e.g. cataclysmic variables, QSOs, and white dwarfs; {\tt Mix}) but could not be trivially separated using XP spectra alone, and sources to which we could not assign a particular astrophysical label ({\tt NA}). Figure~\ref{fig:specific_user_labels} illustrates an example of the NN classifications for different \msl with an extended version for the entire set of {\tt user\_labels} shown in Fig.~\ref{fig:NN_user_label_subplot}.

\subsection{Selection of likely \Ha emitters \label{subsec:selection}}

\begin{figure}
    \centering
\includegraphics[width=0.99\columnwidth]{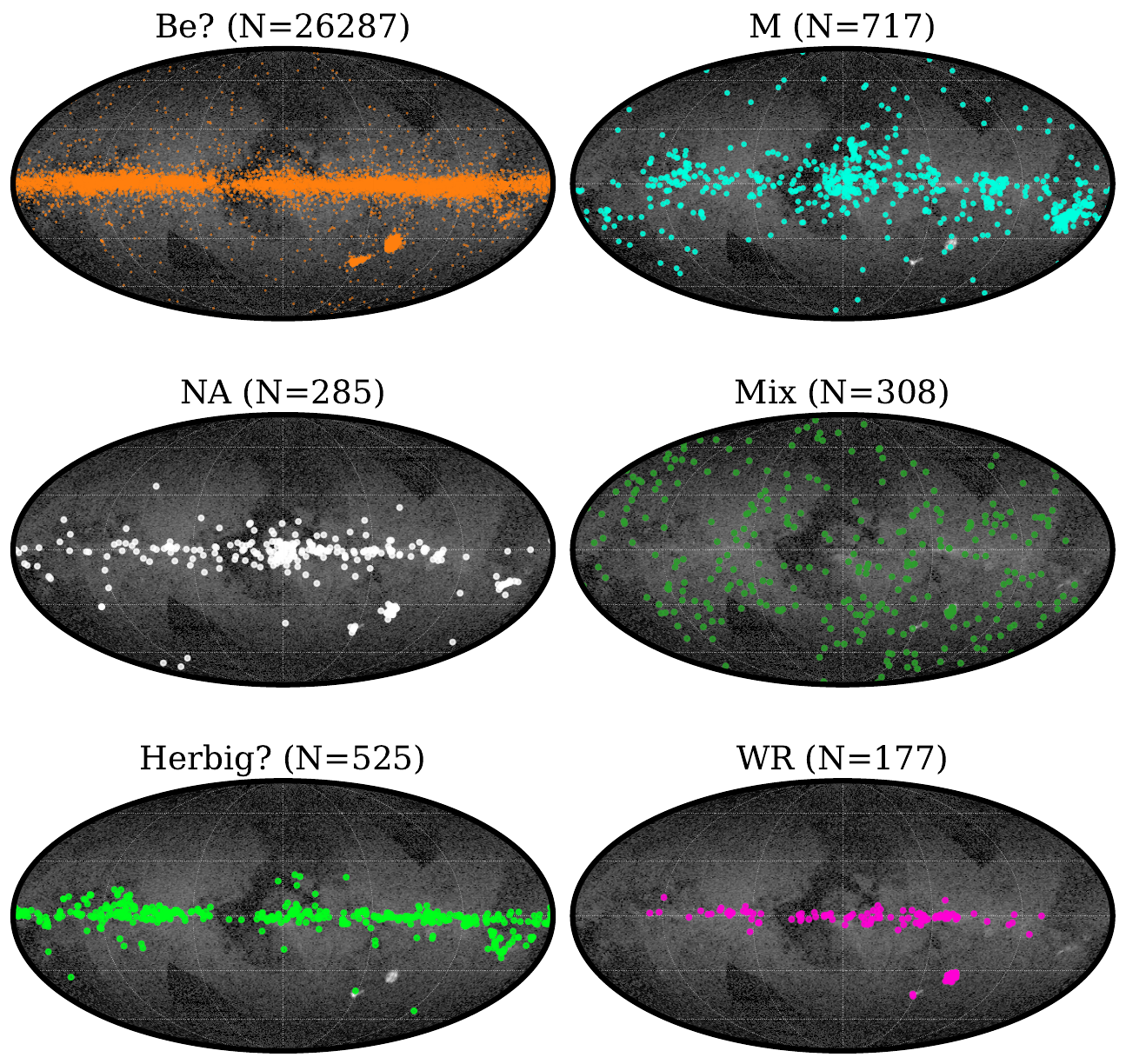}
    \caption{Sky distributions of the main \msl classes of \Ha emitters in our catalogue. The assigned label and number of identified emitters in various classes are provided at the top of each subplot. All 556\,100 sources in the initial sample are shown by grey points in the background.}
    \label{fig:mollweide_emitters_grid}
\end{figure}

As explained in the previous section, the classification of sources based on their mean \Gaia XP spectra allowed us to separate various stellar types, such as distinguishing Be-like XP spectra from cooler M-type stars depicting molecular bands. This facilitated a better selection of \Ha emitters. After an extensive visual inspection of their spectra, we concluded that all sources (i.e. S/N (\Ha) $\geq2$) with \msl~$\in$~({\tt Be?}, {\tt Mix}, {\tt WR}, {\tt Herbig?}, {\tt [WR]}, {\tt PN}) and with \texttt{classification\_flag}~$\neq0$ (i.e. they are not missclassifications) were bona fide \Ha emitters. We note that the blend of helium (\ce{He II} 6560 \AA) and carbon (\ce{C II} 6578, 6582 \AA) lines in Wolf-Rayet stars can overlap with the \Ha region \citep{Torres1987, Marchenko2004, Neugent2012, Hainich2014}, frequently leading to their (mis)classification as \Ha emitters \citep{Hopewell2005, Marin2024, Mulato2025}. Since the emission lines of Wolf-Rayet winds are intrinsically broad, it is therefore difficult to establish whether \Ha emission is genuinely present, particularly at the resolution of \Gaia DR3 XP spectra. We nevertheless retained them in our sample of likely \Ha emitters, while recognising that a significant fraction of {\tt WR}s might not exhibit \Ha emission. For the rest of the sources, we used a threshold on the relative emission to find additional likely emitters. To this end, we computed the ratio of the flux at the \Ha line to the flux at the nearest extremum (typically at $\sim671.8$ nm) detected by XP-TEAL, \texttt{ratio\_1}, and set a threshold of $\texttt{ratio}\_1\geq1.25$ to select likely emission-line sources\footnote{More sophisticated population-specific selection methods can be adopted for different stellar classes, but we used a single threshold here to maintain a simple and homogeneous selection criterion.} with \texttt{classification\_flag}~$=0$ or \msl~$\notin$~({\tt Be?}, {\tt Mix}, {\tt WR}, {\tt Herbig?}, {\tt [WR]}, {\tt PN}, {\tt Blending}\footnote{Sources with the {\tt Blending} label were considered as non-emitters since the resulting spectra can artificially depict an extremum at the \Ha position (also see Fig.~\ref{fig:NN_user_label_subplot}).}). This resulted in the selection of a total of 28\,394 bona fide \Ha emission-line sources, 26\,287 of which had the \msl of {\tt Be?}. We adopted this conservative threshold to prioritise purity over completeness. However, some previously known emitters might be missed due to this constraint because i) the \Ha emission lines in some cool stars are not strong enough to pass the cut, and/or ii) \Ha emitters are often intrinsically variable \citep[e.g.][]{Kovtyukh2011, Barnsley2013, Kumar2023} and the mean \Gaia DR3 XP spectra might not depict a significant emission (see Sect. \ref{sec:followup}). We launched a simple identifier query in SIMBAD to obtain an estimate of the number of previously classified emitters not included in our selection. We found that out of the 23\,874 non-emitters in our sample that were successfully cross-matched with SIMBAD (most of which had \msl $=$ {\tt M}), we lost 124 (0.5\%) sources with emission-line star ({\tt Em*}) as the main object type\footnote{\url{https://simbad.cds.unistra.fr/simbad/sim-fsam}}, and 473 (2\%) sources that were classified as an emission-line star in at least one study. A visual inspection of their XP spectra showed that while some sources do have a bump at the \Ha line position, the strength was not enough to pass the $\texttt{ratio}\_1\geq1.25$ threshold, that is, their mean XP spectra did not show strong enough \Ha emission. This suggests that our selection of emission-line sources does not significantly affect the overall completeness of M-type emission-line stars. Nevertheless, in the catalogue we provide columns \texttt{ratio\_1} and \texttt{ratio\_2} (the latter is the ratio of the flux at \Ha to the flux at the global maximum), enabling users to implement their own filtering schemes for emission-line stars. In Sect.~\ref{sec:cat} we describe the sample of emitters and various stellar populations classified therein.

\section{Gaia-HELIX catalogue: \Gaia \Ha Emission-Line sources Identified using {\tt XP-TEAL}}\label{sec:cat}


\begin{table*}[htbp]
\caption{Summary statistics of the \msl classes and their correspondence to the classification \userlabel.}
\centering
\begin{tabular*}{\textwidth}{@{\extracolsep{\fill}}llll}
\hline\hline
manual\_spectral\_label & user\_label & N (S/N $\geq 3$) & likely\_halpha\_emitters \\
\hline
\noalign{\vskip 4pt}
{\tt M} & [0, 1, 2, 3, 8, 9, 42] & 516279 (152066) & 717 \\
{\tt Be?} & [4, 5, 6, 7, 52, 56, 67] & 26287 (20601) & 26287 \\
{\tt FGK?} & [88] & 11455 (206) & 2 \\
{\tt Herbig?} & [12] & 525 (468) & 525 \\
{\tt Mix} & [30, 44, 45] & 308 (80) & 308 \\
{\tt QSO} & [11, 15, 21, 27, 47, 48, 80] & 204 (81) & 27 \\
{\tt WR} & [13, 14, 20, 43, 51, 60, 68, 73, 82] & 177 (173) & 177 \\
{\tt CV?MB} & [70] & 51 (24) & 23 \\
{\tt C?} & [26, 59, 66] & 18 (18) & 8 \\
{\tt \lbrack WR\rbrack} & [22, 23, 25, 40] & 12 (12) & 12 \\
{\tt Blending} & [31] & 6 (2) & 0 \\
{\tt PN} & [61, 72, 74, 86] & 4 (4) & 4 \\
{\tt NA} & Remaining sources & 583 (349) & 285 \\\\
\texttt{classification\_flag = 0} &  & 191 (0) & 19 \\\hline
\textbf{\textit{Total}} & \textbf{} & \textbf{556100 (174084)} & \textbf{28394} \\
\hline
\end{tabular*}
\label{tab:helix_summary}
\end{table*}

We name the final catalogue of 556\,100 analysed sources containing 28\,394 bona-fide \Ha emitters the {\it Gaia}-HELIX catalogue: \Gaia \Ha Emission-Line sources Identified using {\tt XP-TEAL}. A detailed description of the format and the use of the catalogue is provided in App.~\ref{app:use_cat}, and Tab.~\ref{tab:helix_summary} provides a summary of the classifications and the number of likely \Ha emitters in each class. 

\begin{figure}
    \centering
\includegraphics[width=0.99\columnwidth]{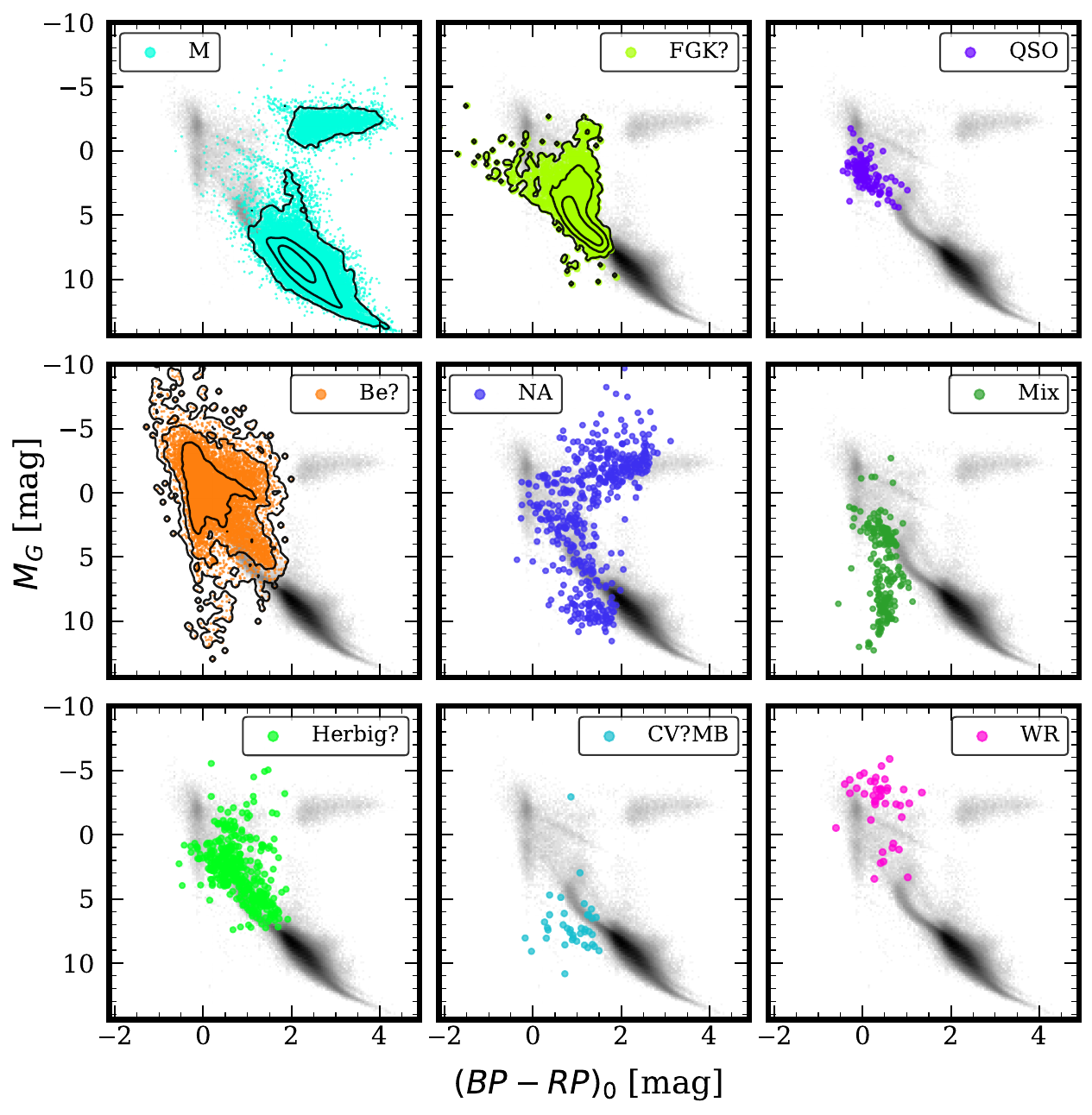}
    \caption{CAMDs constructed using absolute $G$ magnitudes and intrinsic colours for different \msl classes, derived from {\tt SHBoost} \citep{Khalatyan2024} sources with \texttt{xgb\_logteff\_outputflag} = 0. The density contours are overlaid for classes containing more than 2000 sources. The top left panel ({\tt M} stars) contains mostly non-emitters (see Table \ref{tab:helix_summary}). }
    \label{fig:camd_msl_shboost}
\end{figure}

Figure~\ref{fig:mollweide_emitters_grid} shows the sky distributions of the selected emitters grouped by their \msl. This figure already gives us important clues about the astrophysical origin as well as the purity and completeness of these different classes. For example, the Be (class {\tt Be?}), Herbig~Ae/Be (class {\tt Herbig?}), and Wolf-Rayet star (class {\tt WR}) candidates are mostly confined to the Galactic plane and the Magellanic Clouds, as expected for young stars with masses~$\gtrsim2M_{\odot}$. On the other hand, the {\tt M} stars with emission lines are much more scattered in Galactic latitude, since they are expected to be located much closer to the Sun due to observational constraints. The sky distribution of the {\tt Mix} sample appears almost random, with only the imprint of the \Gaia DR3 scanning law visible (see Fig.~\ref{fig:scanning_law_snr}). This suggests that this mixture of objects might be dominated either by very local objects or by spectra with spurious instrumental effects.

\begin{figure}
    \centering
\includegraphics[width=0.99\columnwidth]{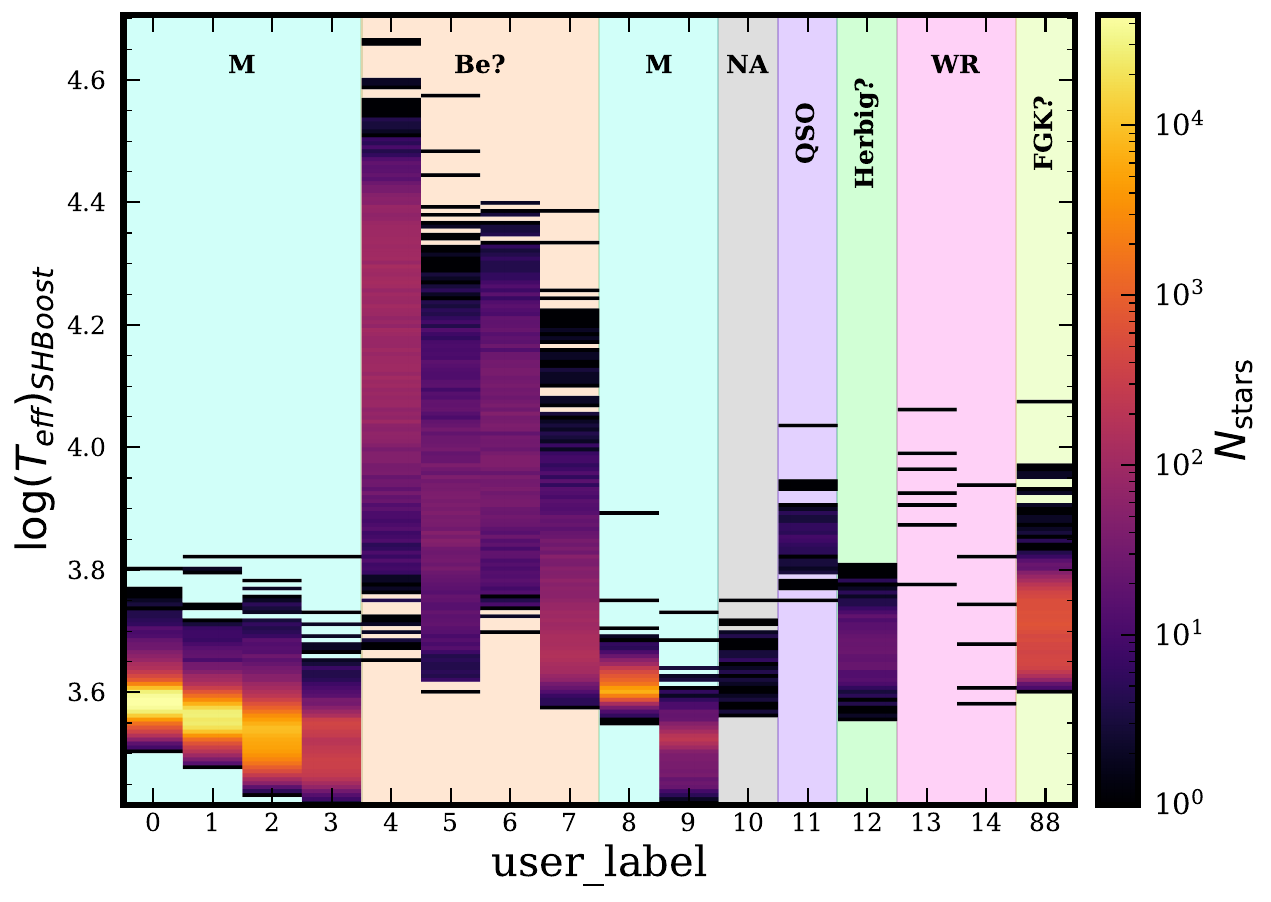}
    \caption{{\tt SHBoost} effective stellar temperatures for some of the most populated NN classification labels, with the shaded regions indicating the corresponding {\tt manual\_spectral\_labels}.}
    \label{fig:SHBoost_teff_vs_user_label}
\end{figure}

For further illustration, Fig.~\ref{fig:camd_msl_shboost} shows the deredenned colour versus absolute magnitude diagrams (CAMDs) for the nine most populated {\tt manual\_spectral\_labels} using the extinction corrected\footnote{\referee{\citet{Khalatyan2024} used \href{https://github.com/fjaellet/gaia_edr3_photutils}{\texttt{gaia\_edr3\_photutils}} to derive the extinction in \Gaia EDR3 passbands given {\tt SHBoost} extinctions and effective temperatures.}} \referee{colour ({\tt bprp0}) and absolute magnitude ({\tt mg0})} from {\tt SHBoost} \citep[Tab. B.1]{Khalatyan2024}. \referee{The stellar} parameters in {\tt SHBoost} were obtained by training an {\tt XGBoost} regressor on \Gaia DR3 XP spectra, photometry, and parallaxes, and stellar labels (effective temperature, surface gravity, interstellar extinction, metallicity, and stellar mass) mostly obtained from Bayesian isochrone fitting of stars contained in spectroscopic stellar surveys with {\tt StarHorse} \citep{Queiroz2023}. The CAMDs reinforce the classification scheme we presented in Sect.~\ref{subsec:nn_classification}. The most populous \msl classes {\tt M} and {\tt Be?}, as well as {\tt FGK?}, {\tt Herbig?}, {\tt CV?MB}, and {\tt WR} are generally found in the regions of the CAMD expected for their respective stellar populations. The diagrams also provide insight into the possible astrophysical nature of objects assigned to the uncertain classes {\tt Mix} and {\tt NA}. Nevertheless, a small number of sources occupy locations in the CAMD that appear inconsistent with their assigned classifications. For example, several objects classified as {\tt Be?} are located in regions typically associated with white dwarfs. We discuss these discrepancies and their implications in more detail in Sect.~\ref{subsec:limitations}.

Nevertheless, we caution that these CAMDs were constructed from distances, effective temperatures, and interstellar extinctions that were based on standard stellar evolutionary models, while many \Ha emitters are products of non-standard single or binary stellar evolution \citep{Kogure2007}. Therefore, we do not expect the {\tt SHBoost} CAMDs of these objects to be completely trustworthy. An extreme example is the case of the {\tt QSO} class, which contains a high fraction of previously known QSOs (Sect.~\ref{subsec:qso} and Tab.~\ref{tab:lit_comparison}) where the CAMD of this class was constructed using distances that were computed under the assumption that they were stars within the Milky Way, hence rendering them unphysical for objects outside the Galaxy.

In Fig.~\ref{fig:SHBoost_teff_vs_user_label} we plot the {\tt SHBoost} effective temperature distributions for several {\tt user\_labels}, demonstrating that our classifier reliably groups stars by their $T_{\rm eff}$, at least in the regimes in which the {\tt SHBoost} $T_{\rm eff}$ values are reliable ($\lesssim 10\,000$ K; see \citealt{Khalatyan2024}). Moreover, different {\tt user\_labels} within one particular \msl can correspond to different spectral subtypes, for instance for the labels {\tt M} and {\tt Be?}. In conjunction with SIMBAD classifications, Figs.~\ref{fig:mollweide_emitters_grid} through \ref{fig:SHBoost_teff_vs_user_label} helped us to associate the {\tt user\_labels} with astrophysical object classes.

\subsection{Classical Be stars}\label{subsec:cat_be_stars}

The XP spectra of most of our newly found \Ha emitters (Tab.~\ref{tab:helix_summary}) are compatible with classical Be stars, that is, B-type main-sequence stars with strong Balmer emission lines (see Fig.~\ref{fig:specific_user_labels}). First identified in the nineteenth century \citep{Secchi1866}, Be stars are characterised by the presence of circumstellar material, typically in the form of equatorial discs \citep{Struve1931} that produce infrared (IR) excess through free-free and free-bound emission \citep{Gehrz1974}. They are also known for their rapid rotation and pronounced variability \citep[e.g.][]{Slettebak1979, Porter2003, Rivinius2013, Rivinius2026}. Their formation is often attributed to binary interactions, although recent observations have challenged scenarios requiring a high multiplicity fraction \citep{Kalari2025}. Nevertheless, many Be stars are found in binary systems, including systems with compact or otherwise unseen companions, and some exhibit strong X-ray emission \citep[e.g.][]{Reig2011, Fedorova2025}. In addition, some Be stars have been proposed as potential surviving companions ejected during Type~Ia supernova explosions \citep{Bao2026}.

\begin{figure}
    \centering
\includegraphics[width=0.99\columnwidth]{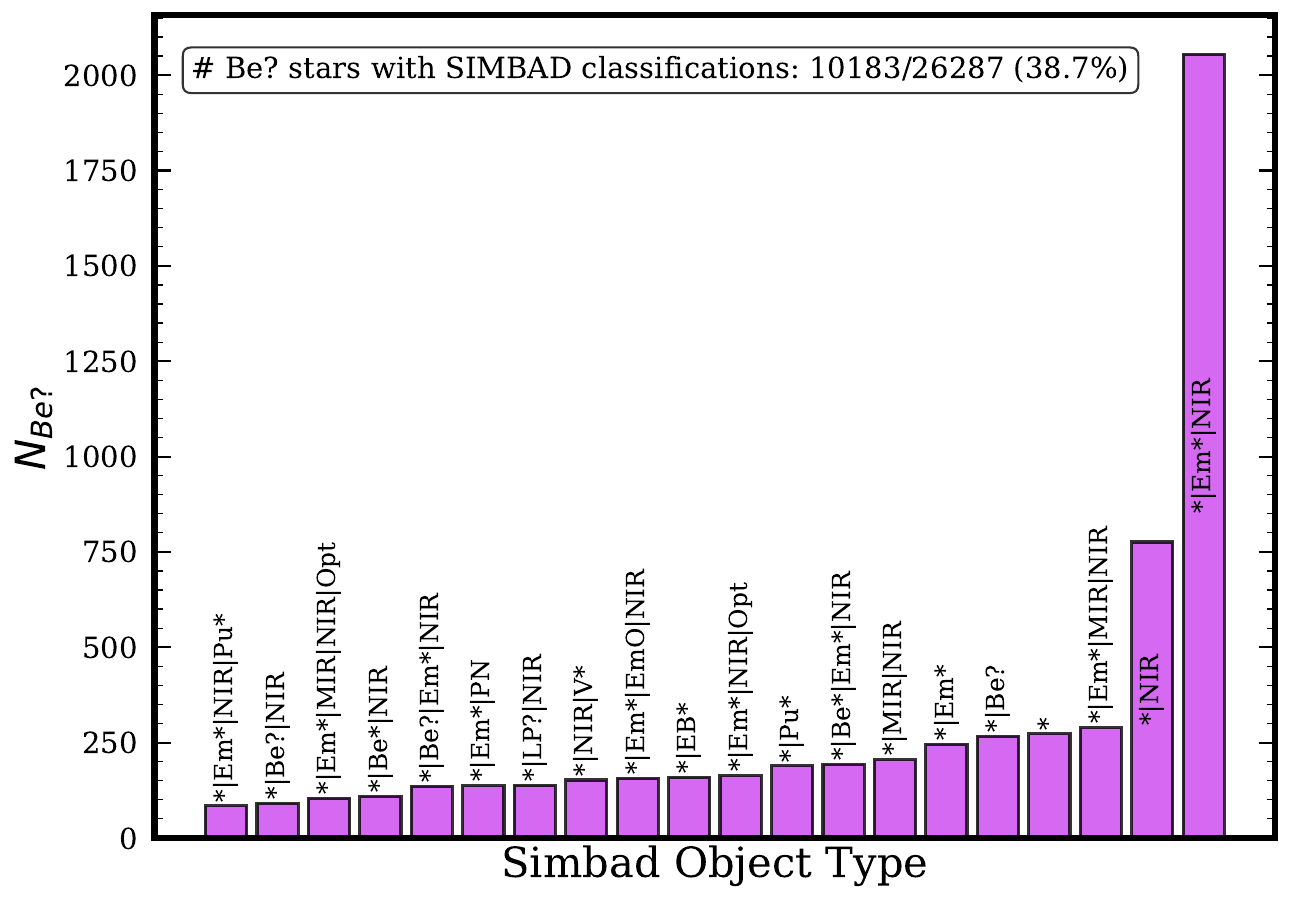}
    \caption{Distribution of available SIMBAD classifications for 10\,183 {\tt Be?} sources.}
    \label{fig:be_candidate_simbad_otype}
\end{figure}

Figure~\ref{fig:be_candidate_simbad_otype} shows the classifications from SIMBAD for the stars we flag as Be-like XP spectra ({\tt Be?}) in our catalogue. The SIMBAD labels were available for 10\,183 out of 26\,287 {\tt Be?} sources. Thus, most of the cross-matched stars appear as emission-line stars with near- or mid-IR excess in SIMBAD, which inspires further confidence in our classification of these sources (the \Gaia XP spectra do not extend to the IR). Furthermore, since A-type stars exhibit the strongest Balmer absorption lines, we expect negligible contamination from Ae sources when \Gaia low-resolution XP spectra alone are used for the classification. This is supported by our comparison with SIMBAD classifications, where only 0.58\% of sources (60 sources) with available classifications were identified in at least one study as {\tt Ae*} or {\tt Ae?}.

\subsection{Herbig Ae/Be stars \label{sec:herbig}}

Herbig~Ae/Be stars (named after \citealt{Herbig1960}) are intermediate-mass ($2$--$15\,{\rm M}_\odot$) pre-main-sequence objects \citep{Strom1972} that are still in the contraction phase and are typically associated with circumstellar dust, such as accretion discs \citep{Waters1998, Malfait1998}. These stars represent an important bridge between low-mass stars ($\mathrm{M} \lesssim 1.5\,\mathrm{M}_\odot$ and massive ($\mathrm{M} \gtrsim 10\,\mathrm{M}_\odot$) stars, and they are also used to investigate planet formation processes (see \citealt{Brittain2023} for a recent review). In addition to strong \Ha emission, Herbig stars frequently exhibit \ce{Ca II} near-IR triplet emission lines \citep{Hamann1990, Carmona2010, Nidhi2023}.

While it is often challenging to distinguish Herbig stars from T Tauri stars (their lower-mass counterparts, i.e. pre-main-sequence precursors of F-, G-, K-, and M-type stars; \citealt{Hamann1992, Vink2005, Valegard2021}) and classical Be stars \citep{Rivinius2013}, several studies (e.g. \citealt{Finkenzeller1984, Vioque2020, Shridharan2021}) have used the stronger IR excess in Herbig stars arising from thermal emission from circumstellar dust to better separate them from Be stars. On the other hand, \citet{Agundez2018} examined the disc chemical composition to distinguish them from T~Tauri stars. This ambiguity, particularly when relying on low-resolution optical spectra alone, is also reflected in the comparison with other \Gaia XP-based classifications in Sect.~\ref{subsec:classification_comp}, although a rigorous identification is beyond the scope of this work.

\begin{figure}
    \centering
\includegraphics[width=0.99\columnwidth]{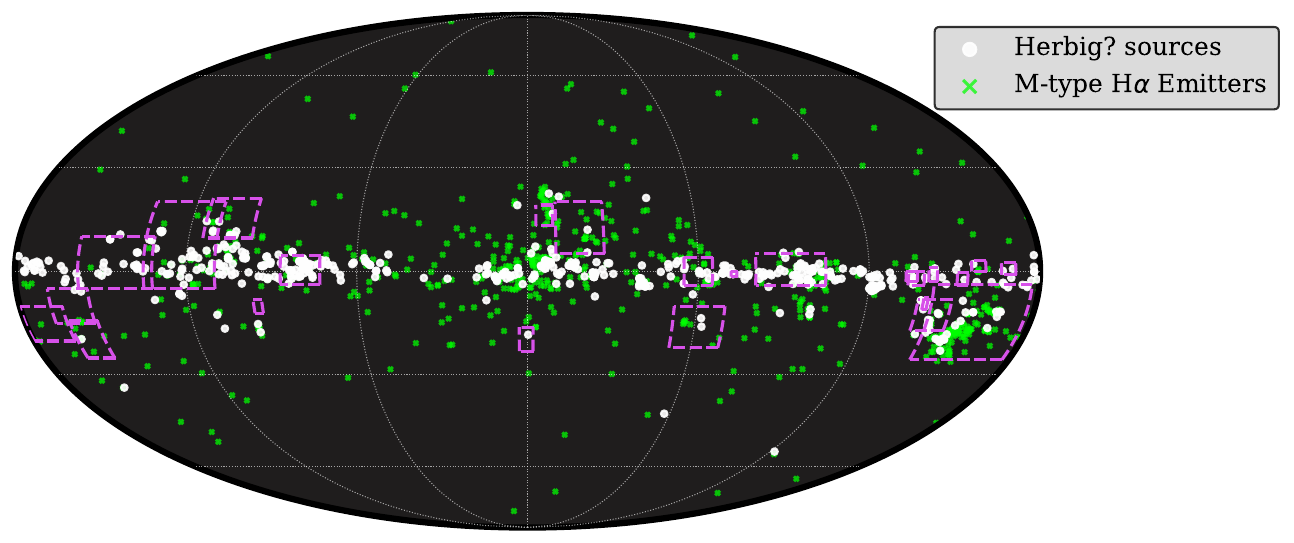}
    \caption{On-sky distribution of Herbig~Ae/Be ({\tt Herbig?}) and \referee{M-type emitters}, together with the approximate positions of prominent nearby molecular clouds (marked as violet rectangles; see text).
    }
    \label{fig:herbig_mollweide}
\end{figure}

We show the on-sky distribution of Herbig~Ae/Be candidates together with the approximate locations of several well-known molecular clouds \citep{Zucker2019, Dharmawardena2023} in Fig.~\ref{fig:herbig_mollweide}. Most of the identified Herbig candidates are concentrated in regions close to the Galactic plane and appear to be associated (in the 2D projection) with the molecular clouds.

\subsection{\referee{M-type emission-line stars}}

The objects in this class span a range of \Ha emission mechanisms, although not all are recoverable from \Gaia XP spectra. Chromospheric and coronal activity in low-mass main-sequence stars, driven by strong magnetic fields, flares, and starspots \citep{Bell2012}, produces \Ha equivalent widths of a few \AA\ at most \citep{Kiman2021}, typically below the emission strengths detectable at the resolution of the \Gaia DR3 XP spectra. Indeed, the weakest M-type emitter in our catalogue has EW(\Ha) = 26.1 $\pm$ 12.7 \AA. Active M dwarfs are therefore unlikely to contribute significantly to our sample, except perhaps in the most extreme cases.

The emission we recovered is instead likely to arise predominantly from two channels, both of which produce \Ha emission that is far stronger than chromospheric activity. The first is magnetospheric accretion in pre-main-sequence T~Tauri stars, where material channelled from the circumstellar disc onto the stellar surface gives rise to broad and intense \Ha emission \citep{Kurosawa2006, Lima2010}. These stars are preferentially found in star-forming regions \citep{Joy1945, Herbig1962}, consistent with the spatial distribution of our candidates (Fig.~\ref{fig:herbig_mollweide}), which exhibits overdensities coincident with well-known star-forming regions and molecular clouds.

The second channel is binary interaction in symbiotic systems, where a hot compact companion accretes material from a red giant \citep{Munari2019}. This mechanism likely accounts for the $\sim$110 M-type emission-line stars located on the red giant branch in Fig.~\ref{fig:camd_msl_shboost}, consistent with the 94 of these objects ($\sim$86\%) that were classified as symbiotic-star candidates by \citet{Merc2026}. Shock-induced \Ha emission in pulsating Mira variables can also contribute to our sample, but this emission is strongly dependent on the pulsation phase \citep{Richter2001}. The epoch-averaged nature of the \Gaia DR3 XP spectra is likely to dilute such phase-dependent variability, suggesting that this channel probably contributes little to the M-type \Ha emitters considered in this work.

\subsection{Cataclysmic variables with molecular bands}

Cataclysmic variables (CVs) are binary systems in which a white dwarf accretes material from the donor star, leading to the formation of a hydrogen-rich accretion disc that produces strong Balmer emission lines \citep{Warner1995,Ritter2003}. In a subset of CVs, particularly those with low accretion rates, the late-type donor star can dominate the optical continuum and imprint molecular absorption features such as TiO bands, causing these systems to spectroscopically resemble field M dwarfs \citep{Szkody2011, Breedt2014}. We identified 51 such sources in our catalogue, where 11 out of 30 sources with available SIMBAD classifications were classified as CVs and exhibited molecular-band features in their \Gaia XP spectra, with approximately 50\% of them showing \Ha emission.

\subsection{QSOs}\label{subsec:qso}

Quasi-stellar objects (QSOs) are the luminous nuclei of active galaxies powered by accretion onto supermassive black holes, producing a variety of emission lines including \Ha (e.g. \citealt{Antonucci1993}). We classified a total of 204 sources as {\tt QSO} candidates, of which all except 2 were in the \Gaia DR3 QSO catalogue \citep{GaiaCollaboration_Klioner2022}. The two missing sources have $p_{\rm quasar} > 0.97$ in the \citet{Shi2026} catalogue. While this class of objects appears very clean, we note that several objects with QSO-like spectra appear in some other \msl classes (see Appendix~\ref{app:use_cat}).

\subsection{Wolf-Rayet star candidates \label{sec:wolf-rayet}} 

Wolf-Rayet stars (WRs) are massive stars known to be progenitors of core-collapse supernovae and stellar-mass black holes \citep{Wolf1867, vanderHucht1981, Crowther2007, Shenar2026}. A significant fraction (at least $\sim40\%$) of WR stars are known to reside in binary systems \citep{vanderHucht2001, Dsilva2023}. WR stars are broadly divided into the nitrogen (WN) and carbon (WC) subclasses \citep{Smith1968}, reflecting distinct evolutionary stages of envelope stripping: WN stars retain a hydrogen-burning or helium-burning envelope and display strong nitrogen and helium emission lines, while WC stars have lost their nitrogen-rich layers and exhibit prominent carbon and oxygen lines, placing them at a more advanced stage of mass loss \citep{Crowther2007}. We identified a total of 177 WR stars in our analysis, of which 62 sources in {\tt user\_labels} 14, 60, 68, 73, and 82 seem to be of WC type, whereas 115 {\tt WR} in \userlabel 13, 20, 43, and 51 seem to belong to the subclass WN. A cross-match with SIMBAD resulted in 161 classifications, of which all except 8 PN, 3 suspected long-period variables, and 1 CV have {\tt WR} classification in our catalogue, but all are assigned a WR classification either in the Galactic Wolf-Rayet Catalogue\footnote{\url{http://pacrowther.staff.shef.ac.uk/WRcat/index.php}} \citep{Rosslowe2015} or by the ESP-ELS pipeline. Figure ~\ref{fig:mollweide_emitters_grid} shows that all sources classified as WR are confined to the Magellanic Clouds and the Galactic plane, with none observed towards the Galactic anti-centre. 

\begin{figure}
    \centering
\includegraphics[width=0.99\columnwidth]{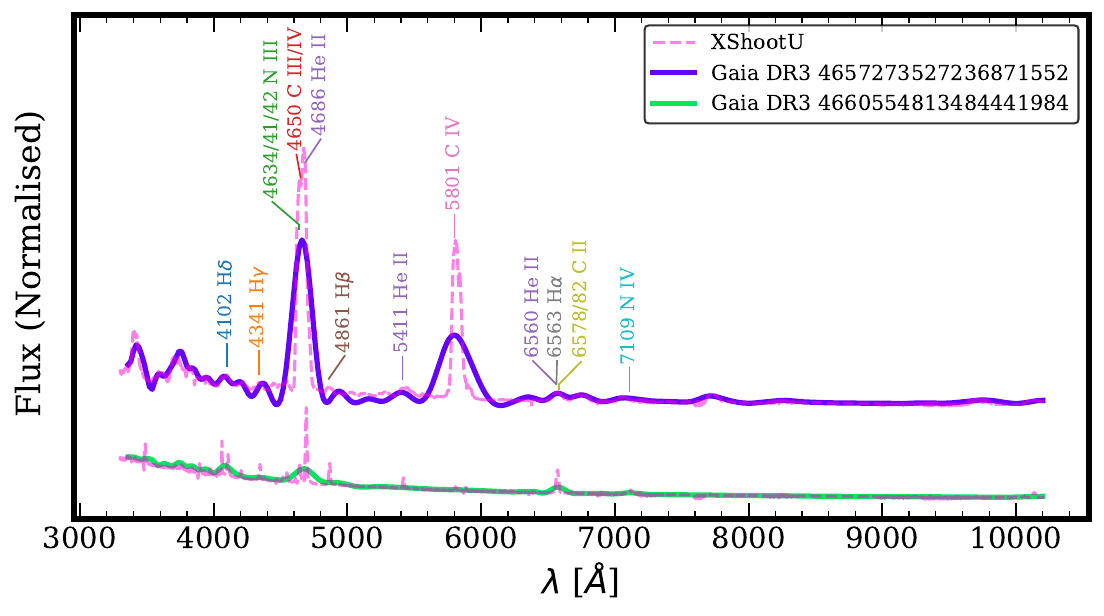}
    \caption{Comparison of externally calibrated \Gaia DR3 XP spectra (solid lines) with higher-resolution spectra (R $\sim$ 9000; dashed pink lines) from XShootU \citep{Sana2024} for two \texttt{WR} stars (\texttt{Gaia DR3 4657273527236871552} and \texttt{Gaia DR3 4660554813484441984}) classified as WC \citep{Torres-Dodgen1988} and WN \citep{Smith1996}, respectively. Some prominent spectral lines in WR spectra are shown by vertical lines.
    }
    \label{fig:wr_spectra}
\end{figure}

Figure~\ref{fig:wr_spectra} shows a comparison between \Gaia DR3 XP spectra and the higher-resolution VLT/X-Shooter spectra from the XShootU project \citep{Vink2023, Sana2024} for two sources (\texttt{Gaia DR3 4657273527236871552} and \texttt{Gaia DR3 4660554813484441984}) identified as \texttt{WR} (\userlabel 14 and 43) in our catalogue. Our NN classifier is clearly able to distinguish between WC and WN sources (the WC shows a very strong broad emission at 5801 $\AA$, which is not seen in WN stars). Moreover, Fig.~\ref{fig:wr_spectra} clearly shows that the blend of \ce{He II} and \ce{C II} coincides with the \Ha line. The broad emission lines hinder a robust characterisation of any underlying \Ha emission, particularly at the resolution of \Gaia DR3 XP spectra. As discussed in Sect.~\ref{subsec:selection}, we retained the 177 {\tt WR} sources in our list of likely \Ha emitters while acknowledging that a fraction of these candidates might not be genuine \Ha emitters.

\subsection{Central stars of planetary nebulae}

The central stars of planetary nebulae that display WR-like spectral features are designated [WR] to distinguish them from massive WR stars, due to their fundamentally different evolutionary origin as low- to intermediate-mass stars in the post-AGB phase \citep{Acker2003,Weidmann2020}. WR and [WR] stars of the same class can depict similar spectra and can usually only be distinguished by using distance information or their location in the HR diagram \citep{Grosdidier2000, Mulato2025}. Since it is not straightforward to separate the two populations using spectra alone \citep{vanderHucht1981, Gorny2001}, we used SIMBAD classifications to flag 12 sources as \texttt{[WR]} candidates.

\subsection{Carbon stars}

Carbon stars (C*) are evolved giants whose atmospheres show a carbon-to-oxygen ratio exceeding unity, producing characteristic molecular absorption bands of \ce{C2} and \ce{CN} in their spectra \citep{Wallerstein1998}. They have been the subject of several recent studies: \citet{Roulston2025} investigated carbon stars in \Gaia DR3, \citet{Ye2025} performed a deep-learning interpretability analysis of C* identification using \Gaia XP spectra, and \citet{Munoz2026} identified 6372 carbon star candidates in LAMOST. Although carbon stars do not typically show \Ha emission, one of the scenarios in which they may be identified as emitters is when they reside in a mass-transferring binary, such as a symbiotic system. The comparison of our catalogue with the symbiotic star catalogue of \citet{Merc2019} in Tab.~\ref{tab:lit_comparison} confirms that 8 of our 18 \texttt{C?} sources are indeed likely to reside in symbiotic systems, of which 4 are identified as \Ha emitters in our catalogue.

\subsection{Unclassified sources}

Owing to degeneracies in the \Gaia XP spectra in different stellar populations, we were unable to assign a unique classification to 583 sources (285 sources were found to be likely \Ha emitters) in our catalogue. In most cases, the corresponding \userlabel is not sufficiently clean and is affected by noisy or ambiguous assignments. For example, \userlabel~17 sources exhibit some of the strongest \Ha emission in our catalogue and show highly similar BP spectra (Fig.~\ref{fig:NN_user_label_subplot}). However, the differences in their RP spectra prevent a robust single-class assignment. A SIMBAD cross-match indicates that while the majority of these objects are planetary nebulae, the sample also includes symbiotic stars. Overall, the most common unclassified sources in our catalogue correspond to the SIMBAD classifications of long-period variables, symbiotic stars, planetary nebulae, emission-line sources, Orion variables, and Be stars.

\subsection{Limitations}\label{subsec:limitations}

We dedicate this section to outlining several limitations of the broad spectral classifications provided in this catalogue. Many of these arise from degeneracies in \Gaia DR3 low-resolution spectra when attempting to distinguish between different stellar types. Additional factors include the absence of complementary observables, such as distance or absolute magnitude, in the classification process, as well as neural network misclassifications and impurities in the training dataset.

For instance, based on \Gaia XP spectra alone, it is not always trivial to distinguish between a reddened B or an AFGK star (e.g. Fig.~26 of \citealt{DeAngeli2023}). Furthermore, the very low resolution of the XP spectra makes it extremely difficult to separate Be stars from cataclysmic variables in cases where the white dwarf dominates the flux (e.g. \citealt{Ambrosch2026}). This limitation is also evident in Fig.~\ref{fig:camd_msl_shboost}, where sources located in the region typically associated with white dwarfs or corresponding unresolved binary systems are predominantly classified as {\tt CV} or {\tt WD} in SIMBAD. As shown in Fig.~\ref{fig:stars_similar_qso_template}, there is also some contamination from QSOs in the \texttt{Be?} label, which were missed when building the training dataset.

Therefore, although comparisons with previously published catalogues in Sect.~\ref{sec:literature} indicate that our custom NN is generally able to robustly classify XP spectra, for example identifying four \userlabel classes (4, 5, 6, and 7) that appear to correspond to Be star–like spectra in different extinction regimes (see Fig.~\ref{fig:NN_user_label_subplot}), we treat these sources primarily as candidates for follow-up studies and caution against interpreting these labels as definitive classifications.

\section{Literature comparison}\label{sec:literature}

\subsection{Classifications}\label{subsec:classification_comp}

\begin{figure}
    \centering
\includegraphics[width=0.99\columnwidth]{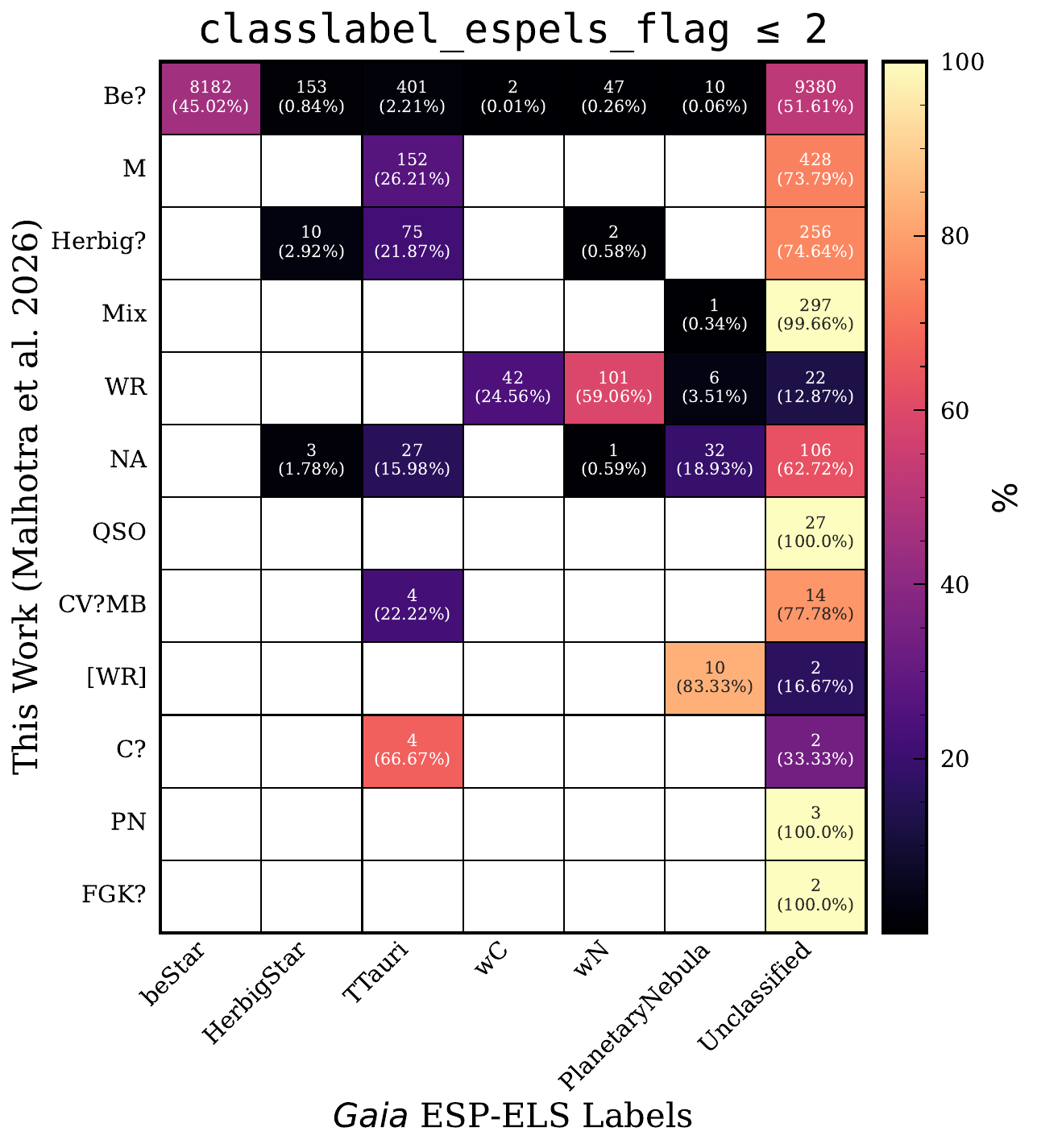}    
    \caption{Comparison of the labels for \Ha emitters in our catalogue to the \Gaia DR3 ESP-ELS labels \citep{Fouesneau2023}.}
    \label{fig:confusion_matrix_manual_spectral_labels_vs_gaiadr3_esp_els_labels}
\end{figure}

One of the largest catalogues of \Ha emission-candidates to date has been produced as part of \Gaia DR3, based on the Extended Stellar Parametrizer for Emission Line Stars (ESP-ELS; \citealt{Fouesneau2023}), one of the modules of {\it Gaia}'s astrophysical parameters inference system ({\it Apsis}; \citealt{Creevey2022}). Figure~\ref{fig:confusion_matrix_manual_spectral_labels_vs_gaiadr3_esp_els_labels} shows a direct comparison between our spectral labels and the classes assigned by the ESP-ELS pipeline. The comparison is not always very favourable, likely owing to differences in class definitions and to misclassifications in ESP-ELS \citep[e.g.][]{Mulato2025} and our catalogue (Sect.~\ref{subsec:limitations}). A relatively good concordance is found for sources classified by ESP-ELS as {\tt beStar}, {\tt wC} and {\tt wN}, and {\tt PlanetaryNebula}. Nevertheless, most of our new {\tt Be?} and emission-line {\tt M} candidates are unclassified in the DR3 ESP-ELS table.

Ideally, we would like to assess the performance (completeness and contamination) of our \Ha emitter selection with higher-resolution observations. Table~\ref{tab:lit_comparison} provides a detailed account of how our catalogue fares when compared with several other \Ha emitter catalogues. We provide the statistics of cross-matches of our classified \Ha emitters with the (mostly recent) literature, covering studies focussing on Be stars \citep{Neiner2011, Chojnowski2017, Vioque2020, An2026, Garcia-Moreno2026}, QSOs \citep{Shi2026}, symbiotic stars \citep{Merc2019, Merc2026}, CVs (\citealt{Hou2020}; \referee{\citealt{Rodriguez2025}}), WRs \citep{Rosslowe2015}, [WR] stars \citep{Acker2003}, carbon stars \citep{Roulston2025}, pre-MS stars \citep{Vioque2020}, massive stars in the SMC \citep{Shenar2024}, the \Gaia Alerts table \citep{Hodgkin2021}, the LMC-SMC young stellar bridge \citep{Scholch2026}, and the \Ha emitter catalogues of \citet{Shridharan2021} and \citet{Tan2025}. The comparisons in Tab.~\ref{tab:lit_comparison} paint a heterogeneous picture, since we compare to variety of studies of different scope, purity, and completeness. Nevertheless, it should be noted that many studies include not only confirmed classifications, but also candidate systems, some of which may ultimately prove to be contaminants, thereby potentially accounting for some of the differences in the comparison table. Depending on their science case, potential users may wish to focus on different aspects of the cross-match table to assess the usefulness of our and previously published catalogues.

\subsection{Comparison of \Ha equivalent widths}\label{subsec:eqwidths}

\begin{figure}
    \centering
\includegraphics[width=0.99\columnwidth]{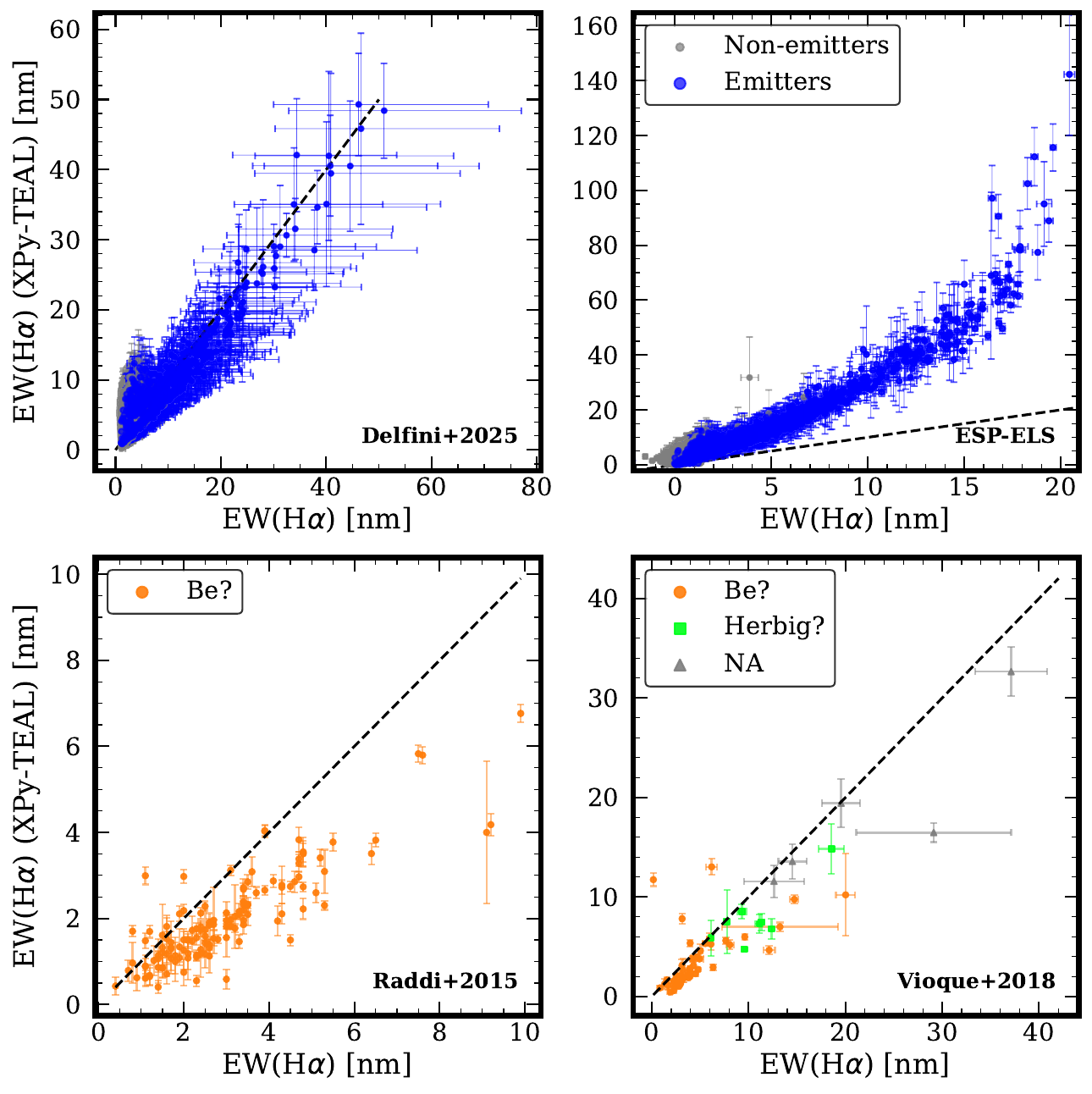}
    \caption{Comparison of our EW(\Ha) measurements with \citet{Delfini2025}, \citet[][ESP-ELS]{Fouesneau2023}, \citet{Raddi2015}, \citet{Vioque2018}. \textit{Top}: Results for emitters and non-emitters identified in this work. \textit{Bottom}: Comparisons for individual stellar populations. The dashed black line indicates the 1:1 relation.}
    \label{fig:eq_width_comparison}
\end{figure}

We compared our \Ha equivalent widths with a small subset of literature estimates, namely \citet{Delfini2025,Fouesneau2023,Raddi2015,Vioque2020}, as shown in Fig.~\ref{fig:eq_width_comparison}. \citet{Delfini2025} used literature-based measurements and applied a linear correction to the pseudo-equivalent widths from \citet{Fouesneau2023}, while \citet{Raddi2015} derived \Ha equivalent widths for Be stars photometrically by interpolating IPHAS colours between theoretical \Ha emission growth curves. In contrast, \citet{Vioque2018} compiled equivalent width estimates from high-resolution spectroscopic studies of Herbig~Ae/Be and Be stars.

Our EW(\Ha) estimates and those of \citet{Delfini2025} and \citet{Vioque2018}, both derived from or calibrated against high-resolution spectroscopy, agree well overall. This consistency is reassuring and supports the robustness of our \Ha equivalent widths from low-resolution spectra. In contrast, we found no comparable agreement with the other studies (as also noted by \citealt{Shridharan2022}), although the consistency is somewhat better for \citet{Raddi2015}.

\subsection{A closer look at Be stars}

\begin{figure}
    \centering
\includegraphics[width=0.99\columnwidth]{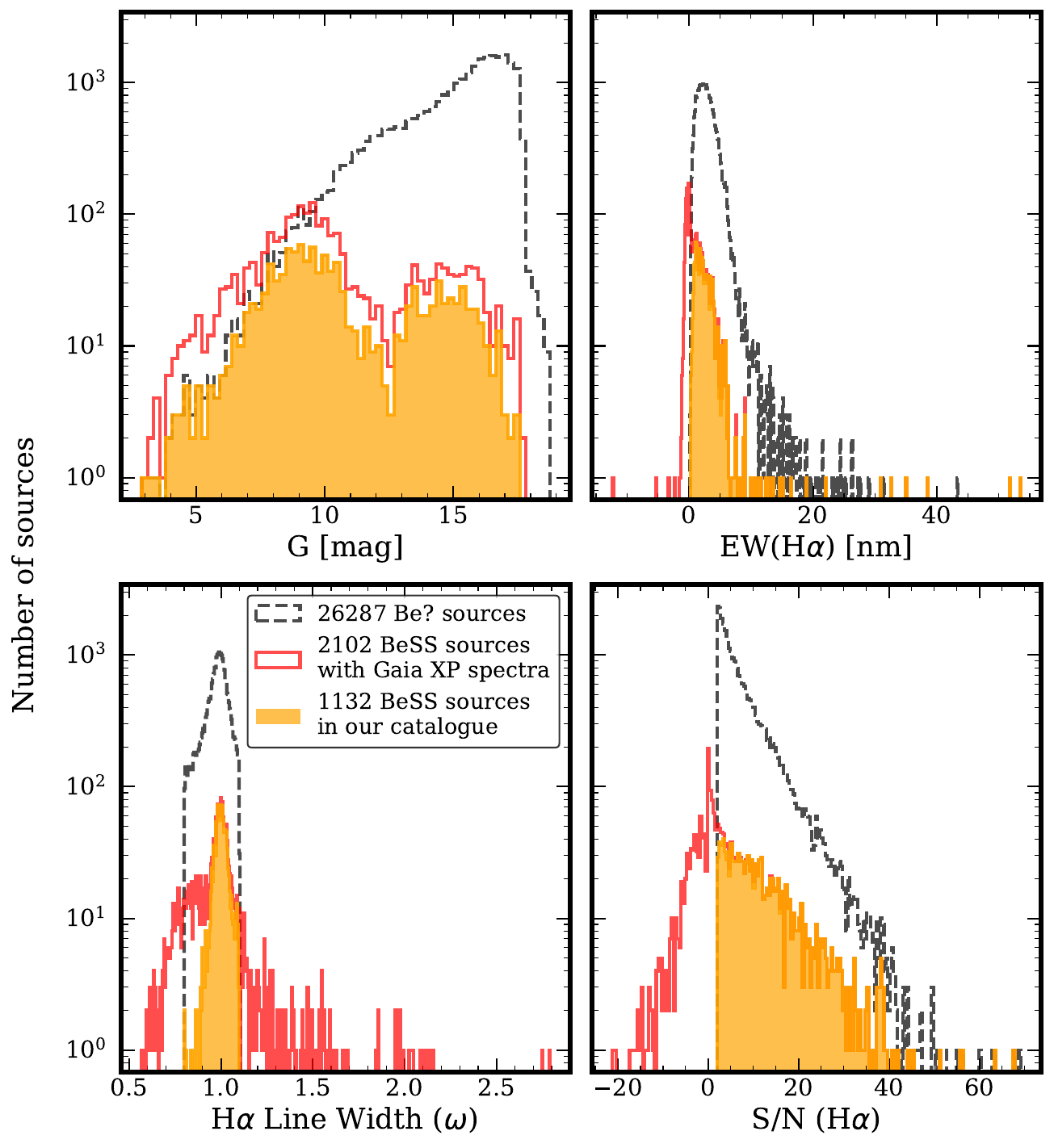}
    \caption{Comparison of our {\tt Be?} sources with the BeSS database. \textit{Top left}: $G$-magnitude distributions. \textit{Top right}: Equivalent width distribution. \textit{Bottom left}: \Ha line width distribution. \textit{Bottom right}: S/N (\Ha) distribution. The number of sources in each of the samples (full {\tt Be?} sample, BeSS sample with DR3 XP spectra, and {\tt Be?}-BeSS crossmatch) is indicated in the legend.
    }
    \label{fig:be_gmag_comparison_bess}
\end{figure}

To study how the constraints we applied to build the initial sample (Sect.~\ref{subsec:xp_teal}) affect the overall completeness of the main component of our catalogue, namely the objects classified as {\tt Be?}, we compared our {\tt Be?} sample with the widely used Be Star Spectra (BeSS) database\footnote{\url{http://basebe.obspm.fr/basebe/}} \citep{Neiner2011}. BeSS objects are not limited to classical Be stars, but their emission has been confirmed, often in multiple observations \citep{Neiner2018}. We plot the distributions of various line parameters and the $G$ magnitude for {\tt Be?} sources while comparing with those of BeSS sources with \Gaia XP spectra and those recovered in our catalogue in Fig.~\ref{fig:be_gmag_comparison_bess}. Although no significant selection bias based on $G$ magnitude is observed, our S/N (\Ha) $\geq 2$ requirement appears to be the most important constraint in determining the recovery of the previously known Be stars in the literature. In other words, we found that we were able to recover about 95\% of the Be stars with S/N (\Ha) $\geq 2$, and about 54\% of sources that we failed to recover had negative estimated \Ha equivalent widths from {\tt XP-TEAL} (i.e. they depicted local minima at the \Ha position in their XP spectra), with the majority of them identified as extrema in the second derivative of the XP spectra. This does not necessarily mean that they never depicted the Be phenomenon and classifications for such sources were entirely incorrect (see \citealt{CastanonEsteban2024} and discussion in Sect.~\ref{sec:followup}). It is possible that these sources were either not emitting in \Ha when observed by \Gaia or are very weak emitters for which the emission is not clearly detectable in the low-resolution and noise-affected mean \Gaia DR3 XP spectra. Such sources are interesting targets to study with the epoch spectra expected to be published with \Gaia DR4\footnote{\url{https://www.cosmos.esa.int/web/gaia/dr4}}.

The high recovery rate of stars in BeSS again shows that our catalogue prioritises purity over completeness. In Sect.~\ref{sec:followup} we describe our follow-up observations of some of the brightest ($G<9.5$) newly discovered Be star candidates (compare the top left panel of Fig.~\ref{fig:be_gmag_comparison_bess}) that confirm $>90\%$ of them.

\citet{Martin2006} and \citet{Boubert2018} specifically discussed Be stars far from the Galactic plane. In this context, we studied the 137 sources with {\tt Be?} \msl that are located at high Galactic latitudes ($|b| > 20$ deg) and found that most of the sources were classified in SIMBAD as cataclysmic variables or QSOs, reflecting some persisting limitations of our training dataset (in the case of QSOs, see also Fig.~\ref{fig:stars_similar_qso_template}) and the aforementioned difficulty in distinguishing B-type stars from WDs (i.e. CVs) and other symbiotic binaries using \Gaia XP spectra alone. We return to the subject of completeness and contamination in Sect.~\ref{sec:followup}.

\section{Spectroscopic follow-up observations}\label{sec:followup}

\subsection{Montsec observations}\label{subsec:montsec}

To test the reliability of our newly found \Ha emitters and their EWs, we observed a representative sample of the brightest newly found \Ha emitters with the ARES medium-resolution spectrograph ($R\sim12,000$) mounted on the robotic 0.8m Joan Oró Telescope\footnote{\url{https://montsec.ieec.cat/en/astronomy/tjo/}} at the Montsec Observatory in Catalonia, Spain. We selected 48 candidates with $G<9.5$ to be observed from Montsec during the first quarter of 2026 (Dec $> 10\,\mathrm{deg}$, \mbox{0\,deg < R.A. < 210\,deg}). We also added three stars that were classified as Be-type runaway stars by \citet{Carretero-Castrillo2023}, but were not found to be reliable emitters in the {\it Gaia} DR3 XP spectra. Of the 21 stars included in \citet{Carretero-Castrillo2023} but not in our list of emission-line objects, only 3 were well observable from Montsec: 1. 
HD 81357 [{\tt Gaia DR3 1026166578637242240}] (an emission-line star according to SIMBAD),
2. BD+61 39 [{\tt Gaia DR3 430511789300158720}] (a Be star according to SIMBAD), and
3. BD+34 113 [{\tt Gaia DR3 362441368008470656}] (an interesting star that was classified as both Be and red giant before).

Our primary goal was to confirm the Be nature of these stars, so we ensured that none of the stars was contained in the BeSS database \citep{Neiner2011}. Their absence from the BeSS catalogue is noteworthy, given that \citet{Neiner2018} claimed that the catalogue is complete down to magnitude $V\simeq11$ mag. Nevertheless, we recall that since Be stars are intrinsically variable on timescales of months to centuries \citep{Barnsley2013, Dimitrov2018}, it is possible that the \Ha emission line appeared and disappeared over time (e.g. during or since the \Gaia observation window).

\begin{figure}
    \centering
\includegraphics[width=0.99\columnwidth, height=11cm]{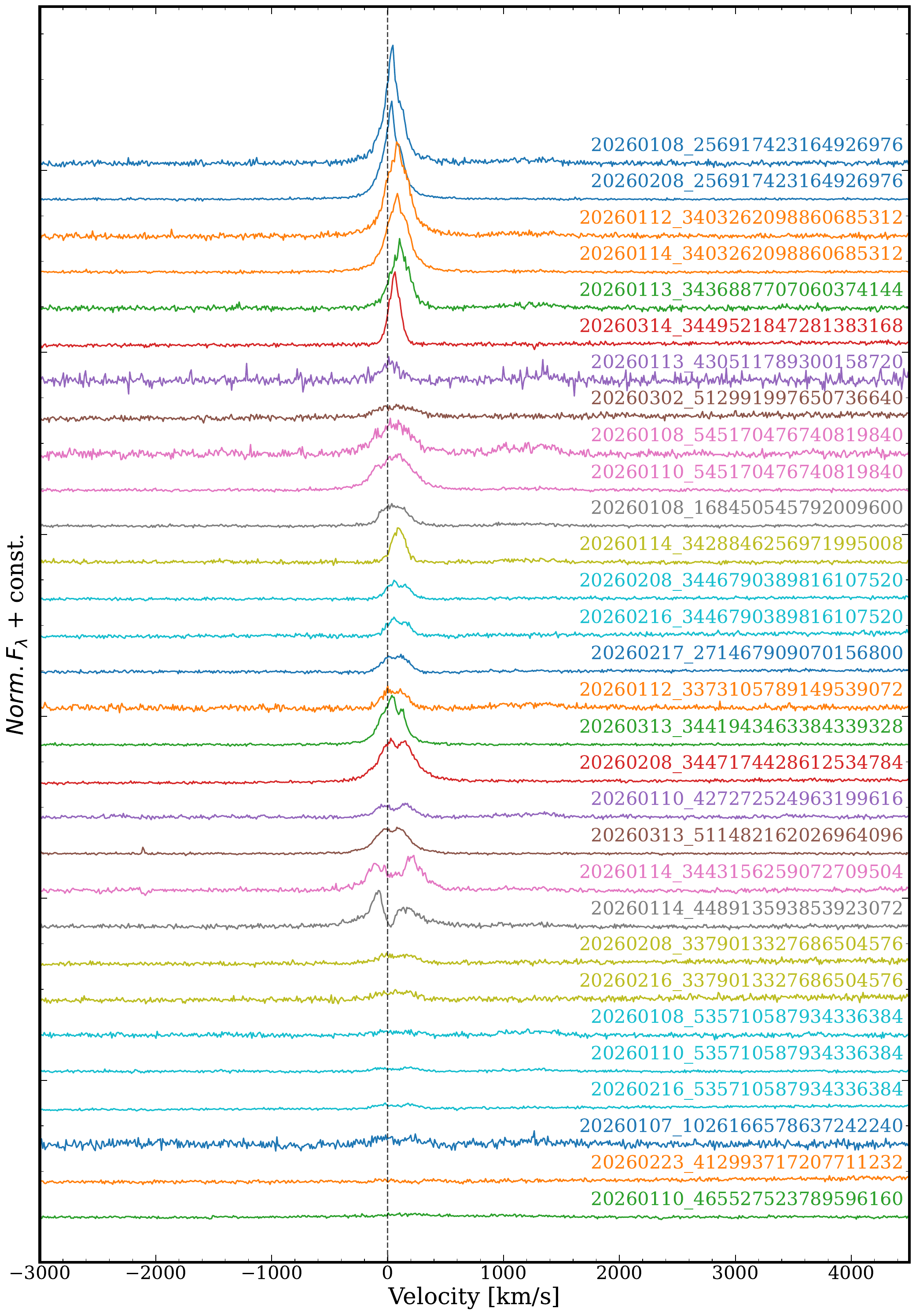}
    \caption{Normalised spectra from 630 to 673 nm of some newly found emission-line candidates (all of them classified as {\tt Be?}) observed with TJO/ARES at Montsec Observatory during semester 2026A, showing a diversity of \Ha line profiles. The adjacent spectra of the same colour correspond to stars that were observed on multiple dates (spectra are labelled as {\tt obsdate\_sourceid}). Low-S/N spectra are shown in Fig.~\ref{fig:tjo_halpha_emitters_low_SNR}.
    }
    \label{fig:tjo_halpha_emitters}
\end{figure}

We finally observed 31 targets (including HD 81357 and BD+61 39 from \citealt{Carretero-Castrillo2023}) successfully (with at least three exposures each) during the 2026A semester. The data were reduced with the standard ARES pipeline. For illustration, we show their co-added spectra in Figs.~\ref{fig:tjo_halpha_emitters} and \ref{fig:tjo_halpha_emitters_low_SNR}. Figure~\ref{fig:tjo_halpha_emitters} shows the normalised ARES spectra of 23 stars with higher-S/N spectra. \Ha is clearly in emission in 19 of them (and 2 possible weak detections in \texttt{Gaia DR3 1026166578637242240} and \texttt{Gaia DR3 535710587934336384}), with a variety of different shapes such as well-defined single peaks to asymmetric double peaks to noisy broad emission, as expected for Be stars. For some of the stars that were observed at several epochs (e.g. \texttt{Gaia DR3 535710587934336384}), variability in the shape of the \Ha line is also indicated. We note that although several other spectral lines fall in the observed spectral window (630 to 673~nm), notably \ce{He I}, \ce{C II}, \ce{Fe II} and forbidden lines such as \ce{[O I]} and \ce{[N II]}, the S/N of the observed spectra is not sufficient to robustly detect or characterise these features if such emission were present.

Most importantly, our observations allowed us to estimate an approximate upper limit of $\lesssim 10\%$ (2 out of 23) for the contamination of the newly found \Ha emitters in the bright regime ($G\lesssim10$). The contamination is probably much lower when we take into account the possibility that the stars without \Ha emission in the ARES spectra might have been emitters during the time they were observed by \Gaia for DR3 (more than 10 years ago). 
Based on long-term monitoring of a sample of Be stars started by \citet{Steele1999}, \citet{Barnsley2013} reported quantitative estimates for the probability that a Be-type star would change state between emission and non-emission. The latest estimate (of $\sim0.75$ per cent per year; \citealt{CastanonEsteban2024}) agrees with our non-detection rate of the new Be candidates, considering the time difference between the {\it Gaia} DR3 and Montsec observations\footnote{Considering a time difference of 10 years, $\sim 0.75\%\times10\times23$, i.e. 1-2 sources are expected to undergo a change between emission and non-emission states.}. We also note that one of the non-confirmed \Ha emitters in the Montsec sample, \texttt{Gaia DR3 412993717207711232}, shows a very prominent \Ha emission line in the DR3 XP spectrum (W(\Ha) $= 1.65 \pm 0.10$~nm; S/N (\Ha) $\sim$ 16). The other source (\texttt{Gaia DR3 465527523789596160})  depicts a relatively weak emission (W(\Ha) $= 0.42 \pm 0.13$~nm; S/N (\Ha) $\sim$ 3) and is classified as a young stellar object in SIMBAD.

In the faint regime, our contamination is likely to be stronger, but we nevertheless recall that the significance of each local maximum associated with a potential \Ha emission is encoded in the rigorously calculated signal-to-noise ratio \citep{weiler2023}, and we only reported \Gaia DR3 sources with S/N (\Ha) $\geq2$ here.

\subsection{GranTeCan and Calar Alto observations}\label{subsec:gtc_caha}

In addition to the Montsec observations, we had three proposals (two with the OSIRIS instrument mounted at the Gran Telescopio Canarias (GTC), using R2500U ($R=2555$, 3440--4610\,\AA{}) and R2500R ($R=2475$, 5575--7685\,\AA{}), with proposal IDs GTC13-25B and GTC13-26A, PI: Garcia-Moreno, G.; and one with the CAFOS instrument mounted at the 2.2\,m telescope at the Calar Alto Observatory, using Grism G--200 ($R \sim 300$, 4000--8500\,\AA{}), with proposal ID: 25B-2.2-016, PI: Garcia-Moreno, G.) for follow-up observations of sources from the second sample in \citep{Garcia-Moreno2026}. This sample was selected by targeting \Ha emitters in the Hertzsprung gap, with variability and mid-IR excess. 

\begin{figure}
    \centering
\includegraphics[width=0.99\columnwidth]{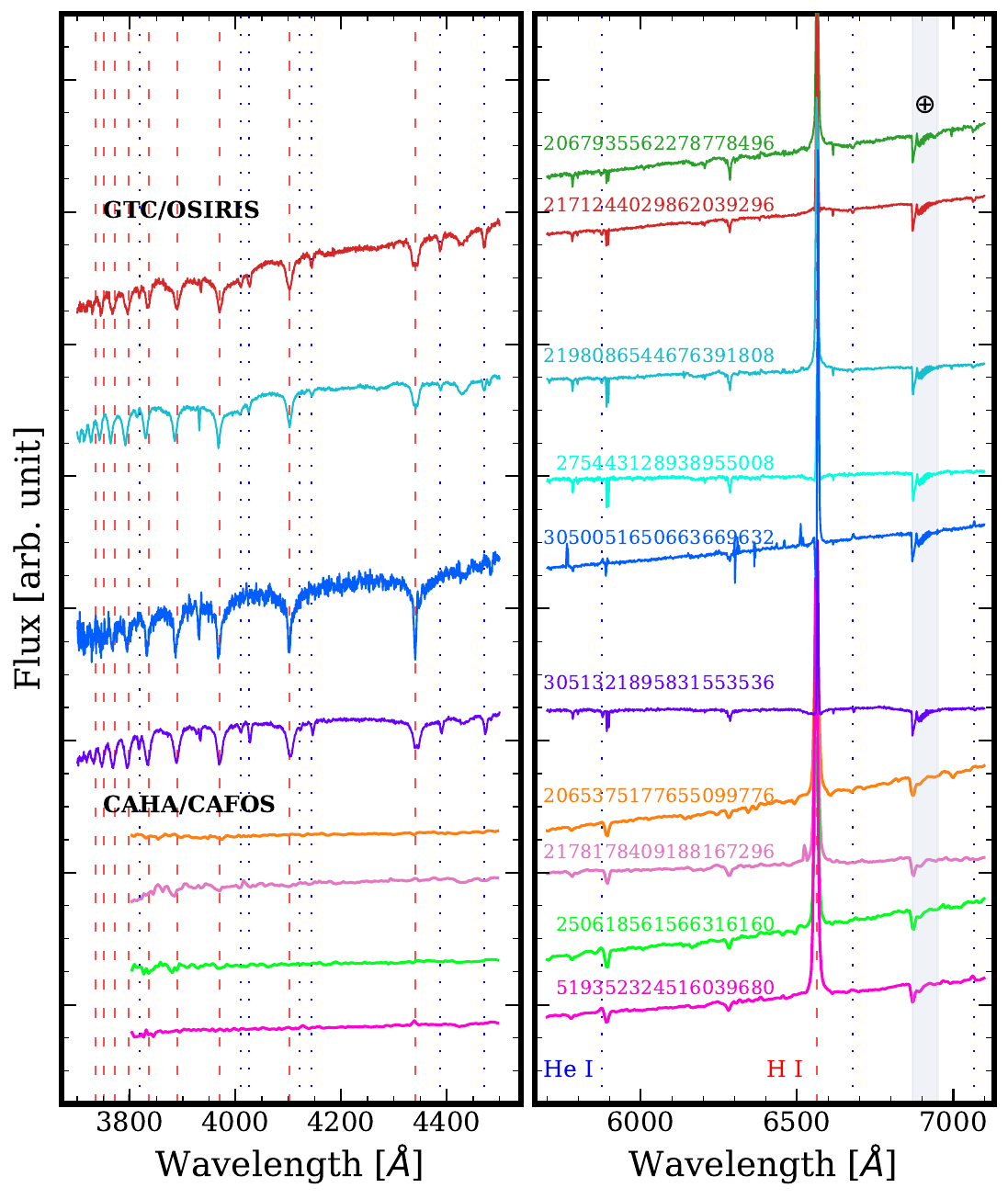}
    \caption{GTC/OSIRIS ($R\sim$ 2500) and CAHA2.2m/CAFOS ($R\sim$ 300) spectra of ten fainter {\tt Be?} sources in our catalogue. $u$-band ($\lambda \sim 3700-4600$ \AA) and $R$-band ($\lambda \sim 5700-7100$ \AA) spectra are shown in the left and right panels, respectively. The crossed circles represent telluric absorption bands, and the dotted blue and dashed red vertical lines indicate \ce{He I} and \ce{H I} lines, respectively.
    }
    \label{fig:gtc_cafos_spec}
\end{figure}

Ten sources matched our catalogue, all of which we classified as {\tt Be?} with user labels 5 or 7. The analysis of the follow-up spectra for these sources revealed that all except one ({\tt Gaia DR3 3050051650663669632}, a very likely Herbig~Ae candidate) are strong candidates for highly reddened Be stars (given their \ce{H I} profiles and the appearance of \ce{He I} absorption lines). This therefore verifies our classification from the NN (Fig.~\ref{fig:NN_user_label_subplot}). We show these spectra in Fig.~\ref{fig:gtc_cafos_spec}. All the observed sources show strong \Ha emission, confirming their emission-line star classification in this work.

These sources were found with the \texttt{SHBoost} catalogue \citep{Khalatyan2024}, which improves extinction estimates for many sources compared to \citet{Anders2022}, but can still be wrong for intrinsic extinction linked to the circumstellar shells around Be stars \citep{Gehrz1974}.

\section{Conclusions \label{sec:conclusions}}

We have carried out a systematic search for \Ha emitters in the full \Gaia DR3 XP dataset, measuring equivalent widths with reliable uncertainties for the Balmer \Ha line at 656~nm. The selection of sources with \Ha emission detected at the $2\sigma$ level or above yielded 556\,100 emission-line candidates. Using a semi-supervised classifier, we provided broad spectral classifications and removed objects that merely masquerade as emission-line stars (mainly cool stars with molecular bands). The resulting Gaia-HELIX catalogue gives a tentative classification for all 556\,100 sources and flags 28\,394 as likely \Ha emitters, the majority with Be-like XP spectra and the remainder spanning M-type emission-line stars, Herbig~Ae/Be candidates, quasi-stellar objects, Wolf-Rayet stars, cataclysmic variables, carbon stars, [WR]-type central stars of planetary nebulae, planetary nebulae, and sources affected by blending, among others.

We foresee several possible applications of our catalogue to further investigations. For example, a cross-match with the classified \Gaia Alerts\footnote{\url{http://gsaweb.ast.cam.ac.uk/alerts/home}} returns 74 matches, most with a \msl of {\tt Be?} or {\tt Mix} and predominantly classified as CVs, QSOs, or YSOs by the Alerts team.

Our literature comparisons, for example with the BeSS catalogue \citep{Neiner2018}, demonstrated that the purity of our catalogue is high ($\geq95\%$). Visual inspection and comparisons with medium- to high-resolution studies showed that our spectral classes (Sect.~\ref{subsec:nn_classification}) are generally consistent given the low resolution of \Gaia XP, largely because the classifier exploits the spectral shape in BP and RP, which is crucial for distinguishing between similar stellar types. Nevertheless, dozens of likely misclassified or doubtful cases may be found, and comparisons to the literature are often plagued by selection biases and small-number statistics.

Finally, we note that infrared and sub-millimeter photometry can substantially improve stellar classification \citep{Malhotra2026}, particularly for pre-main-sequence stars \citep{Ribas2020, Vioque2020}. Our main reason for using only \Gaia DR3 observables in this work was to create a homogeneous catalogue of emission-line sources with a selection function that is not too complex. However, with new IR surveys such as SPHEREx \citep{Melnick2026} and {\it Roman} \citep{Schlieder2024} coming online, it will be worthwhile to include combinations of colours in future studies to further improve our classifications.

\section*{Data availability}

{\tt XP-TEAL} is available at \url{https://xpteal.fqa.ub.edu} with a {\tt python} implementation available at \url{https://github.com/malhotrasagar15/xpy_teal}. Online version of the \Gaia-HELIX catalogue is only available in electronic form at the CDS via anonymous ftp to \href{cdsarc.u-strasbg.fr}{cdsarc.u-strasbg.fr} (130.79.128.5) or via \href{http://cdsweb.u-strasbg.fr/cgi-bin/qcat?J/A+A/}{http://cdsweb.u-strasbg.fr/cgi-bin/qcat?J/A+A/}.

\begin{acknowledgements}

\referee{We thank the anonymous referee for constructive feedback that improved the manuscript.} The authors acknowledge all members of the GaiaUB group for insightful discussions throughout the course of this project. SM acknowledges the grant PRE2022-103025, financed by the Spanish MCIN/AEI/10.13039/501100011033 and the Fondo Social Europeo+. This work was partially supported by "ERDF A way of making Europe" by the “European Union” through grant PID2021-122842OB-C21 and PID2024-1596-1594 of the Institute of OB-C21, and Cosmos Sciences University of Barcelona (ICCUB, Unidad de Excelencia 'Mar\'{\i}a de Maeztu') through grant CEX2024-001451-M and the project 2021-SGR-00679 GRC of the Agency of Management of Universitaris i de Recerca (General Research) of Catalonia. FA acknowledges financial support from MCIN/AEI/10.13039/501100011033 through a RYC2021-031638-I grant co-funded by the European Union NextGenerationEU/PRTR. The Joan Oró Telescope (TJO) at the Montsec Observatory (OdM) is owned by the Catalan Government and operated by the Institute of Space Studies of Catalonia (IEEC). This research was partially funded by the Horizon Europe HORIZON-CL4-2023-SPACE-01-71 SPACIOUS project funded under Grant Agreement no. 101135205. The authors thankfully acknowledge the computer resources from MareNostrum, technical expertise and assistance provided by the Red Española de Supercomputing at the Barcelona Supercomputing Center, National Supercomputing Center. This work has made use of data from the European Space Agency (ESA) mission {\it Gaia} (\url{https://www.cosmos.esa.int/gaia}), processed by the {\it Gaia} Data Processing and Analysis Consortium (DPAC, \url{https://www.cosmos.esa.int/web/gaia/dpac/consortium}). Funding for the DPAC has been provided by national institutions, in particular the institutions participating in the {\it Gaia} Multilateral Agreement. This research has made use of the SIMBAD database, operated at CDS, Strasbourg, France and available at \url{http://simbad.cds.unistra.fr/simbad/}.

\end{acknowledgements}

\bibliographystyle{aa} 
\bibliography{aa61499-26}

\begin{appendix}

\section{Using the catalogue}\label{app:use_cat}

To reproduce samples of sources used in this work, users will need four columns from the catalogue available at the CDS. Those four columns are:
\begin{enumerate}
    \item {\tt likely\_halpha\_emitter}: provides a flag if a source is classified as a likely \Ha emitter
    \item \msl: manually assigned spectral label
    \item \userlabel: output from the neural network
    \item {\tt classification\_flag}: provides a flag if a source is not an obvious misclassification from the neural network. This column only has a value for sources where S/N (\Ha)~$\in$~[2, 3) since higher S/N sources were used as a training dataset for the NN.
\end{enumerate}

In this section we provide the constraints that one can apply to retrieve some specific samples used in the analysis.

\subsection{Query examples}

\textbf{Retrieving likely \Ha emitters}: 
Sources classified as likely \Ha emitters can be obtained by {\tt likely\_halpha\_emitter}~$=1$, leading to a total of 28394 sources.

\noindent
\textbf{Querying by stellar populations}: 
Sources with ${\tt classification\_flag} = 0$ indicate misclassification by the NN, specifically for sources with S/N (\Ha)~$\in$~[2, 3). We recommend always using the constraint of ${\tt classification\_flag} \neq 0$ while querying specific stellar populations. Hence, to retrieve sources classified as M-type, one can use:
\[{\tt manual\_spectral\_label} = {\tt M} \;\land\;{\tt classification\_flag} \neq 0\]
The full list of all manually assigned stellar labels can be found in Tab.~\ref{tab:helix_summary}.

\noindent
\textbf{Querying by \userlabel i.e based on the output of the NN}:
Samples containing sources with the same NN classification can be obtained using the column \userlabel in conjunction with the constraint on the {\tt classification\_flag}. For e.g. sources in \userlabel\xspace{\tt 4} that are Be-like sources with very low line-of-sight extinction (Fig.~\ref{fig:NN_user_label_subplot}) and mostly occupy the region of the Magellanic Clouds can be retrieved as:

\[{\tt user\_label} = {\tt 4} \;\land\;{\tt classification\_flag} \neq 0.\]

\subsection{Cues to removing contaminants}\label{app:remove_contaminants}

In this work, we focus exclusively on classifications derived from the \Gaia XP spectra to construct a homogeneous catalogue of emission-line stars, and we deliberately avoid incorporating additional information such as parallaxes as distance indicators in order to maximise completeness and minimise selection biases. However, as noted in the paper, the purity of individual classes can be further improved when targeting specific stellar populations for detailed follow-up studies. Several auxiliary columns, e.g. {\tt ratio\_1} (ratio of the flux at \Ha line to the flux at the nearest extremum) and {\tt ratio\_2} (ratio of the flux at \Ha line to the flux at global maximum), are provided to allow users to impose additional constraints and thereby improve the purity of likely \Ha emitters within a given class.

As discussed in Sect.~\ref{subsec:cat_be_stars}, a subset of {\tt Be?} sources occupy the white-dwarf region of the HR diagram and are predominantly classified as CVs or WDs in SIMBAD. For these objects, a simple cut on the absolute $G$ magnitude (in particular for \userlabel~4) can help remove such contaminants.

\begin{figure}
    \centering
\includegraphics[width=0.99\columnwidth]{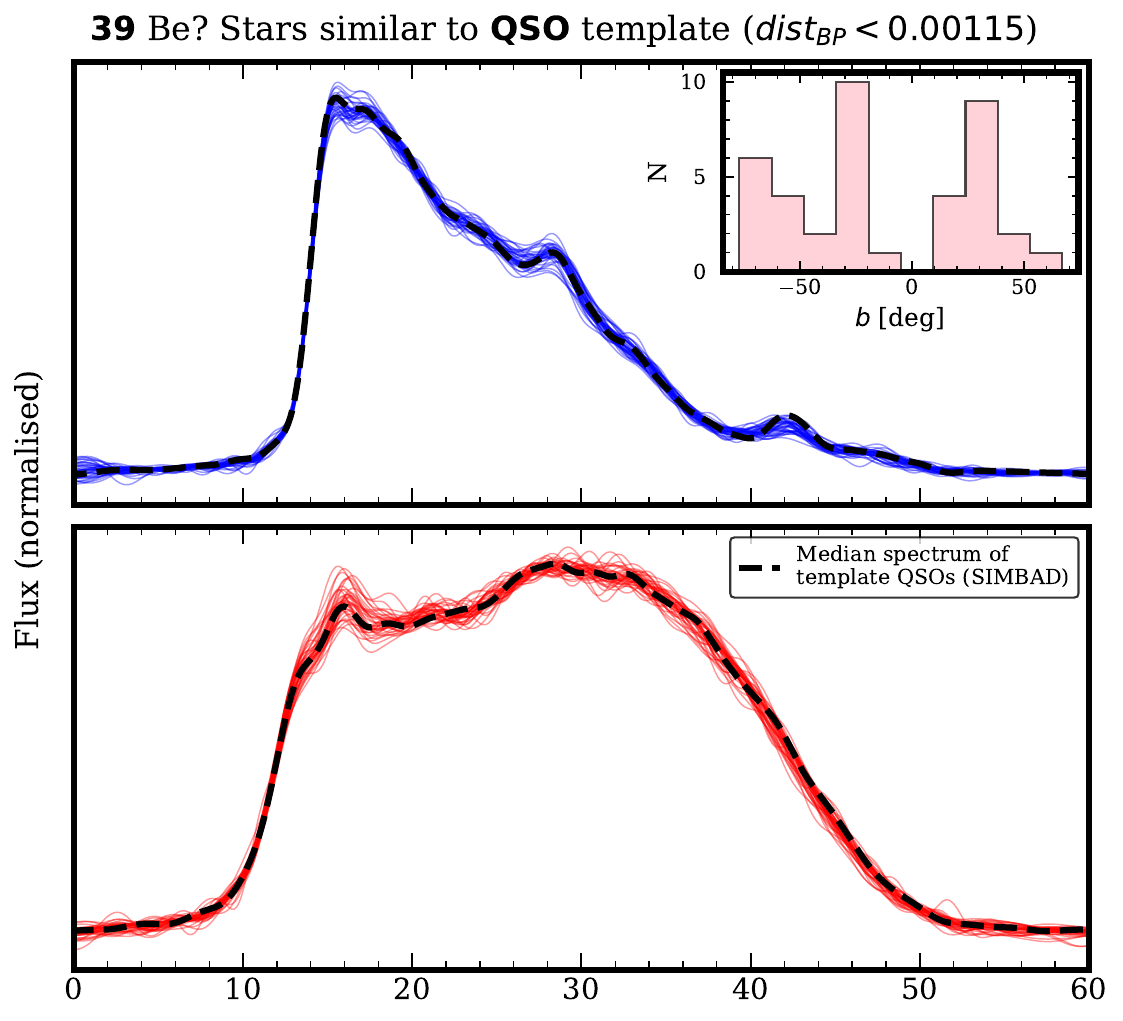}
    \caption{Example of removing specific contaminants from a spectral label. \Gaia DR3 XP spectra of 39 stars identified as contaminants in the {\tt Be?} \msl, selected for their similarity to the median spectrum (shown by the dashed black line) of a sample of QSOs (SIMBAD classifications). \textit{Inset}: Galactic latitude distribution of these sources, showing that most lie away from the Galactic Plane.}
    \label{fig:stars_similar_qso_template}
\end{figure}

Furthermore, in Fig.~\ref{fig:stars_similar_qso_template}, we identify likely QSOs within the {\tt Be?} sample by selecting sources whose BP/RP spectra are similar to the median spectrum of a set of template QSOs, chosen based on their SIMBAD classifications. By applying a suitable distance threshold using the {\tt cosine} metric in the BP band, we identify 39 such contaminant sources within the {\tt Be?} sample. This illustrates how template-matching techniques, based on similarity in mean XP spectra, can be used to remove specific contaminants from a given spectral class.

\section{Using the neural network to classify similar \Gaia DR3 XP spectra}\label{app:NN_features}

\begin{figure*}
    \centering
\includegraphics[width=2.0\columnwidth]{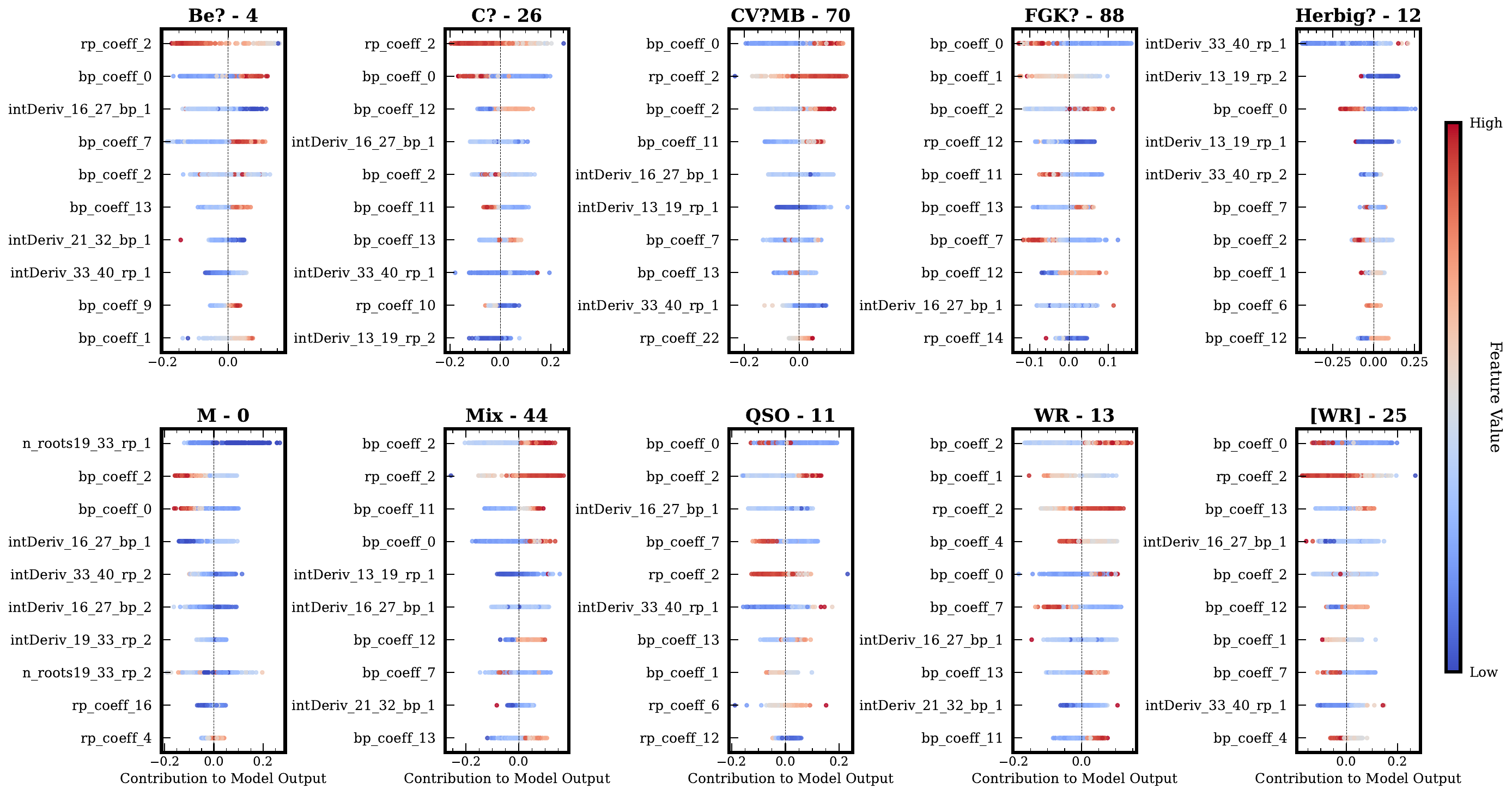}
    \caption{Feature importance for the most populated \userlabel for each \msl.}
    \label{fig:shap}
\end{figure*}

As mentioned in Sect.~\ref{subsec:nn_classification}, we define additional features to capture the shape of the XP spectra, complementing the raw BP/RP coefficients and improving the NN's ability to distinguish between spectral classes. Specifically, we use (i) the number of roots of the $n^{\mathrm{th}}$ derivative of the spectrum within a given pseudo-wavelength range and (ii) the integral of the absolute value of the $n^{\mathrm{th}}$ \referee{[n $\in \{1,2,3\}$]} derivative over the same range. These are denoted as \verb|n_roots{pwl_min}_{pwl_max}_{band}_{n}| and \verb|intDeriv_{pwl_min}_{pwl_max}_{band}_{n}|, respectively. Here, \texttt{pwl\_min} and \texttt{pwl\_max} define the limits of the pseudo-wavelength interval over which the features are computed, and \texttt{band} specifies the BP or RP spectrum, while $n$ indicates the derivative order.

\begin{figure}
    \centering
\includegraphics[width=0.99\columnwidth]{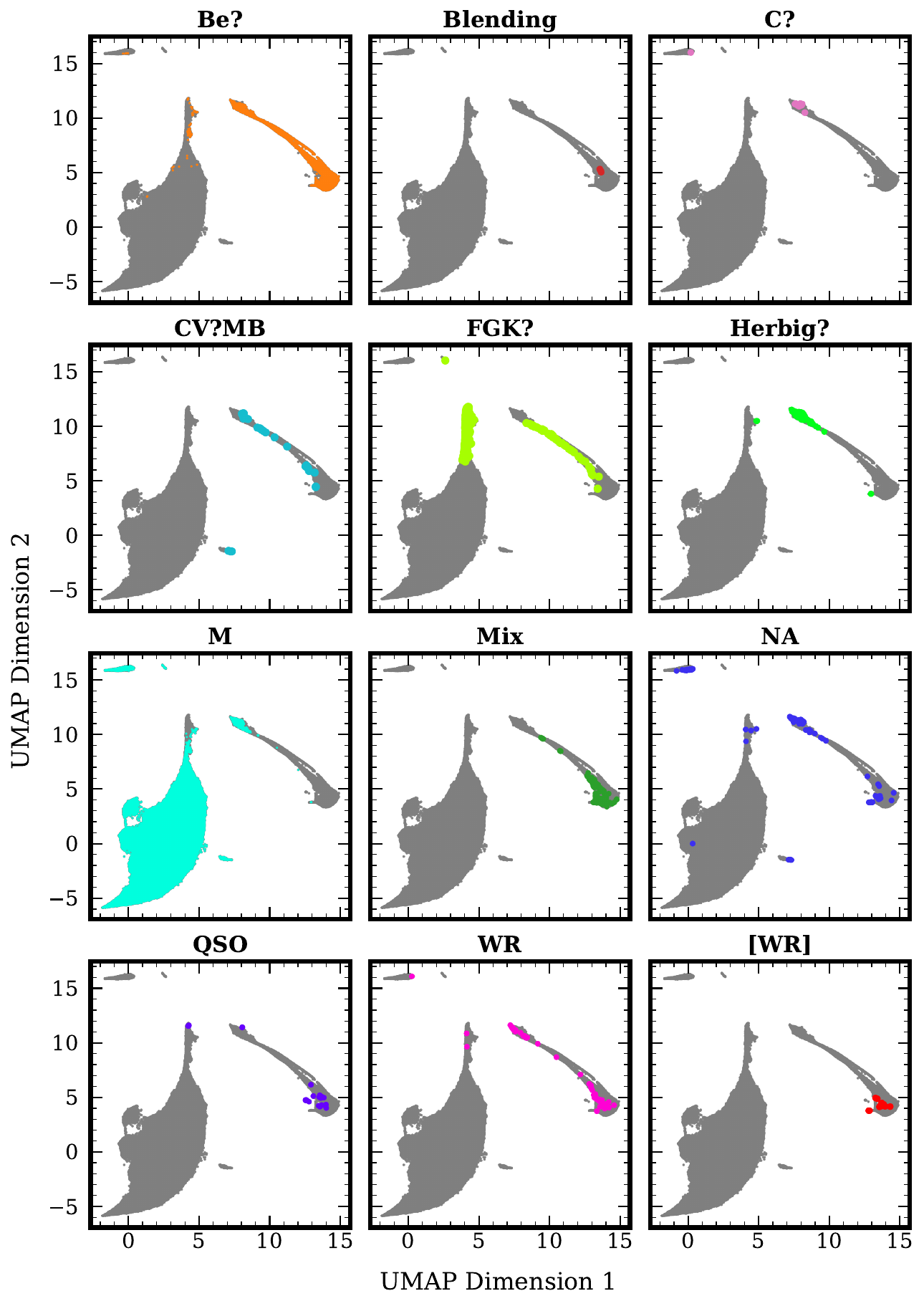}
    \caption{{\tt umap} projection of different stellar types classified in this catalogue.}
    \label{fig:umap_proj_manual_spectral_label}
\end{figure}

For a better interpretability of the classifications of sources based on the XP spectra by the NN, we use the concept of Shapley additive explanations (SHAP\footnote{\url{https://shap.readthedocs.io/en/latest/}}; \citealt{Lundberg2017}). In Fig.~\ref{fig:shap} we plot the SHAP summary plots depicting the 10 most important features for the most populated \userlabel for each \msl and their contribution to the model prediction for a random sample of 1000 sources. In this work, SHAP values are transformed into a normalized quantity that represents the fractional contribution of each feature to the model's decision. Hence, positive values indicate that the feature increases the likelihood of a source being classified to a particular \userlabel with larger positive values implying stronger support. On the other hand, negative values suggest that the feature favours alternative classifications of the source. For example, it is the integral of the absolute value of the \referee{first derivative of} RP spectrum computed within the pseudo-wavelength range of 33 and 40 \referee{({\tt intDeriv\_33\_40\_rp\_1}}; which also encompasses the wavelength range corresponding to the \ce{Ca II} triplet) that is the most important while classifying a source as a {\tt Herbig?} star. Furthermore, for {\tt M} type stars (particularly \userlabel 0 that depicts molecular bands; see Fig.~\ref{fig:NN_user_label_subplot}), it is the number of roots in the \referee{first derivative of} RP spectrum in the pwl range 19 to 33 \referee{({\tt n\_roots19\_33\_rp\_1})}, which becomes significantly important for the model to classify a source into the \userlabel 0.

Another common way to visualise high-dimensional data is through dimensionality reduction techniques such as principal component analysis (PCA), t-distributed stochastic neighbour embedding ({\tt t-SNE}; \citealt{vanderMaaten2008}), and uniform manifold approximation and projection ({\tt UMAP}; \citealt{McInnes2018}). Figure~\ref{fig:umap_proj_manual_spectral_label} shows the {\tt UMAP} projection of our catalogue for each \msl, illustrating how sources with the same label tend to cluster in the 2D embedding. The right-hand cluster is dominated by emission-line sources, while the left-hand cluster consists mainly of cool M-type stars. {\tt FGK?} sources are distributed in both regions, suggesting a possible dependence on stellar mass in the {\tt UMAP} space.

\begin{figure*}
    \centering
    \includegraphics[width=\textwidth]{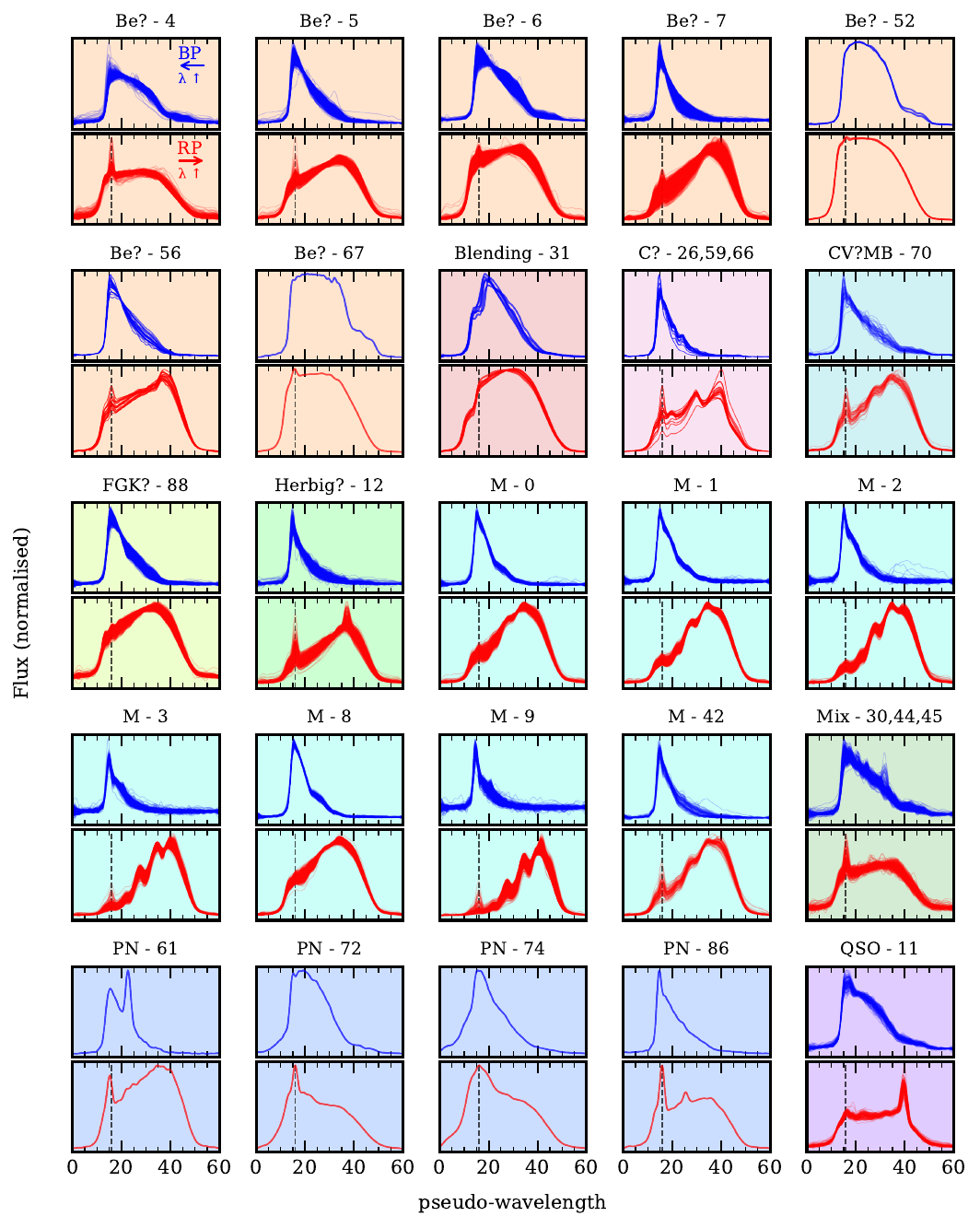}
    \caption{NN classifications: Each subplot (indicated by \msl\xspace - \userlabel) represents a random sample of max. 1000 sources in each \userlabel, containing S/N $\geq$ 3 and S/N $\in$ [2, 3), while classes with the same \msl have the same background colour for the subplot. \Ha line position is indicated by the dashed black line. Sources with {\tt classification\_flag} = 0 are not shown in the figure.}
    \label{fig:NN_user_label_subplot}
\end{figure*}

\begin{figure*}\ContinuedFloat
    \centering
    \includegraphics[width=\textwidth]{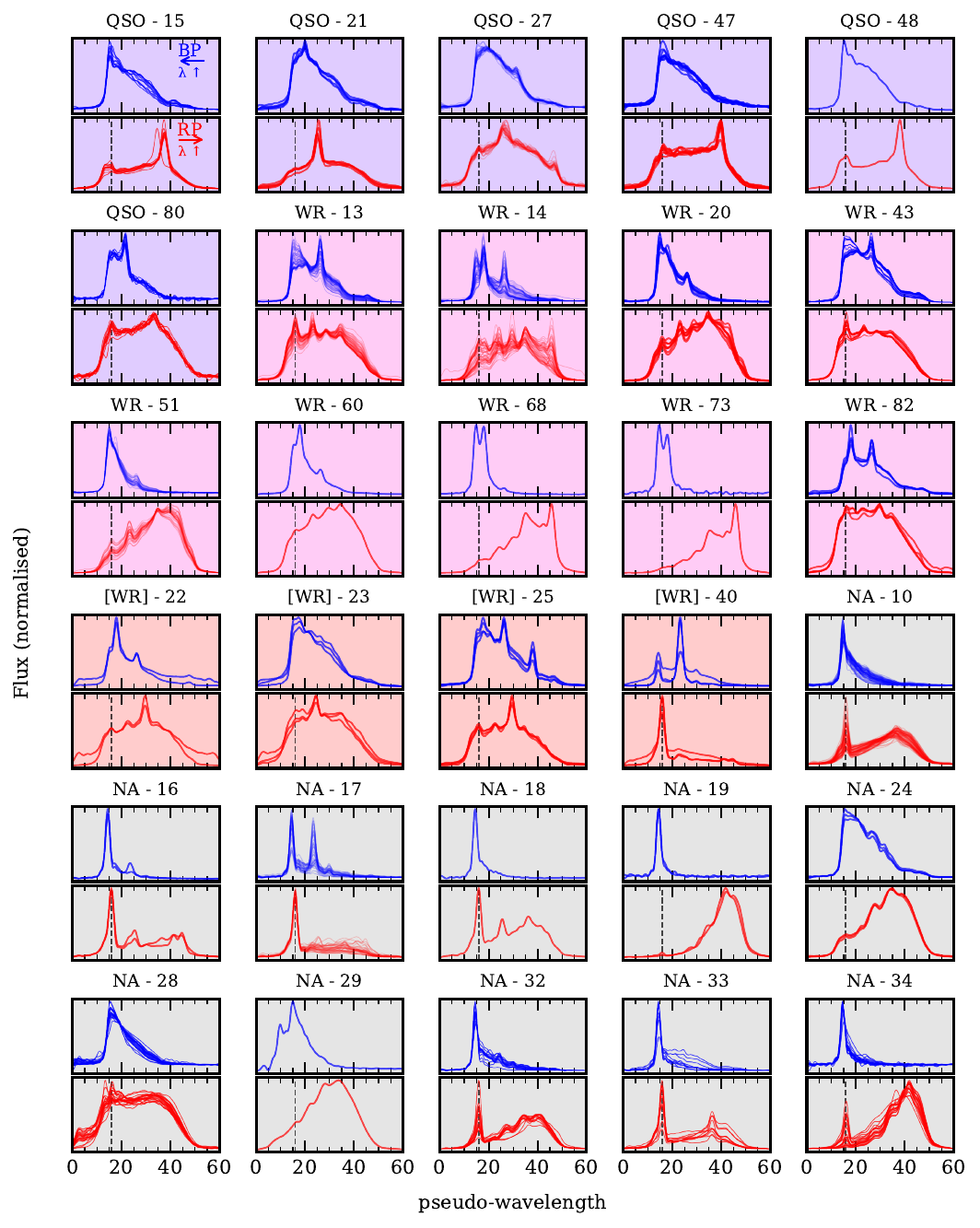}
    \caption{Continued.}
\end{figure*}

\begin{figure*}\ContinuedFloat
    \centering
    \includegraphics[width=\textwidth]{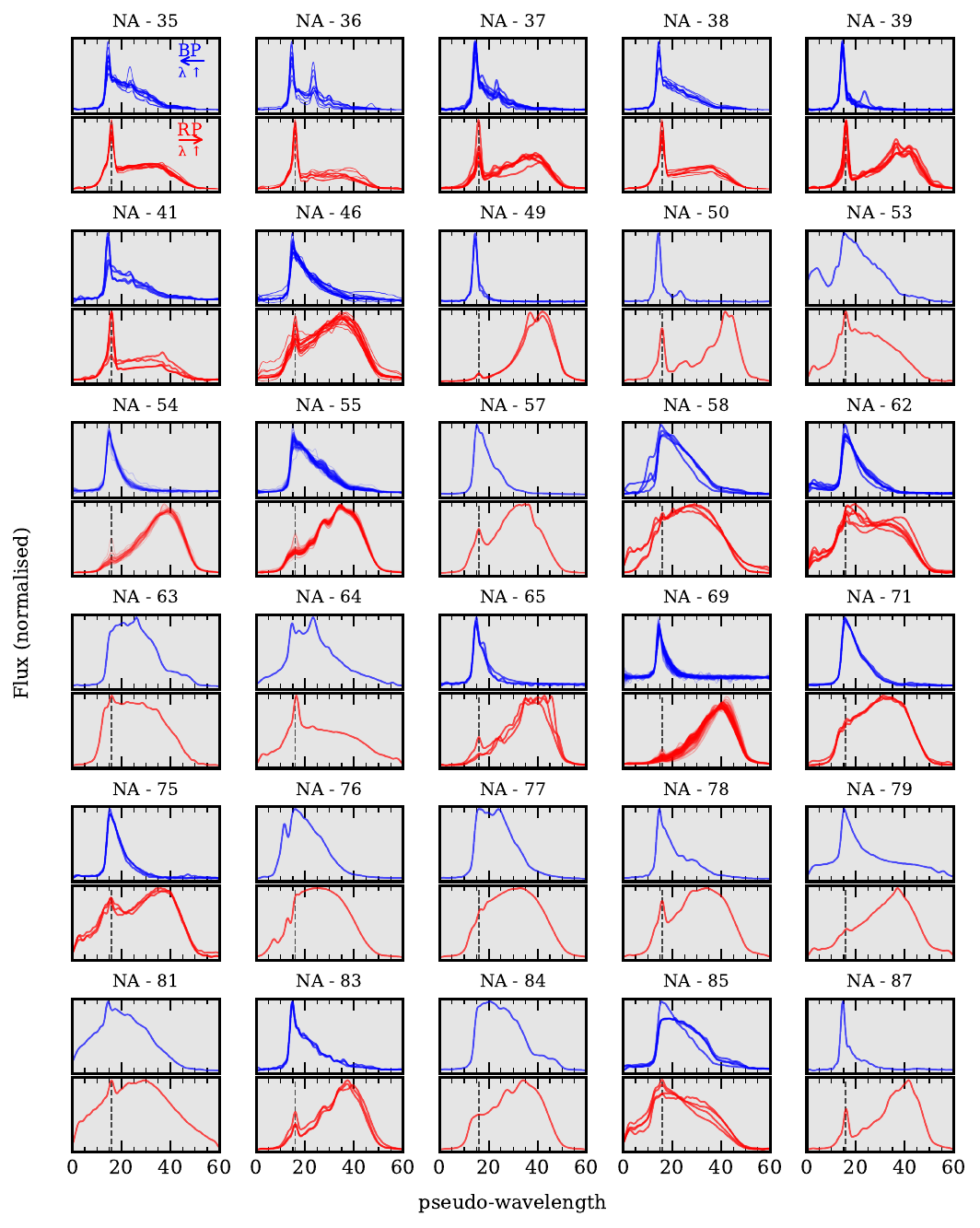}
    \caption{Continued.}
\end{figure*}

\section{Additional figures and tables}\label{app:add_figs_tables}

\begin{figure}
    \centering
\includegraphics[width=0.95\columnwidth]{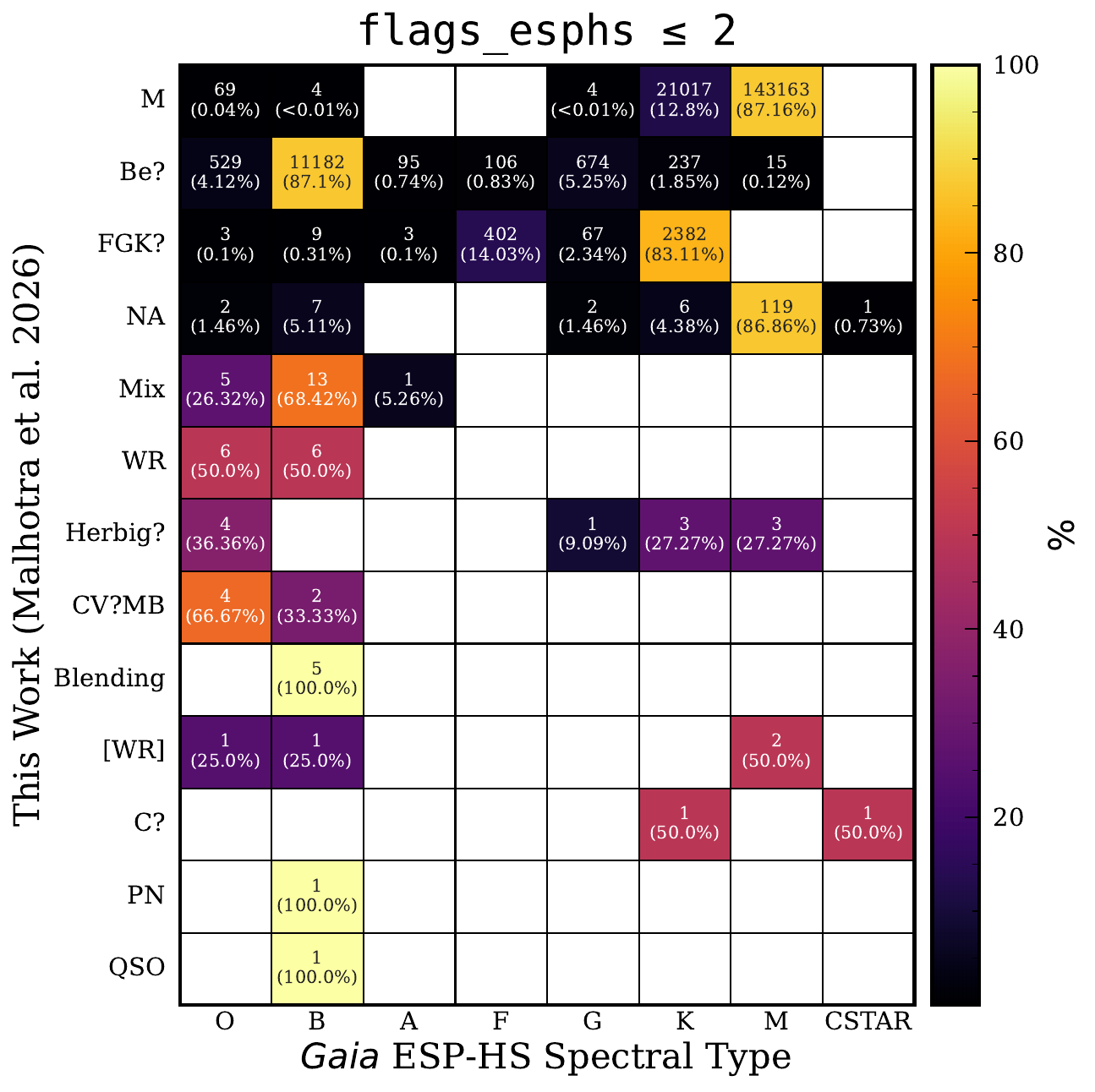}    
    \caption{Comparison of the labels in our catalogue to the \Gaia DR3 ESP-HS catalogue labels \citep{Fouesneau2023}.}
    \label{fig:confusion_matrix_manual_spectral_labels_vs_gaiadr3_esp_hs_labels}
\end{figure}

\begin{figure}
    \centering
\includegraphics[width=0.95\columnwidth]{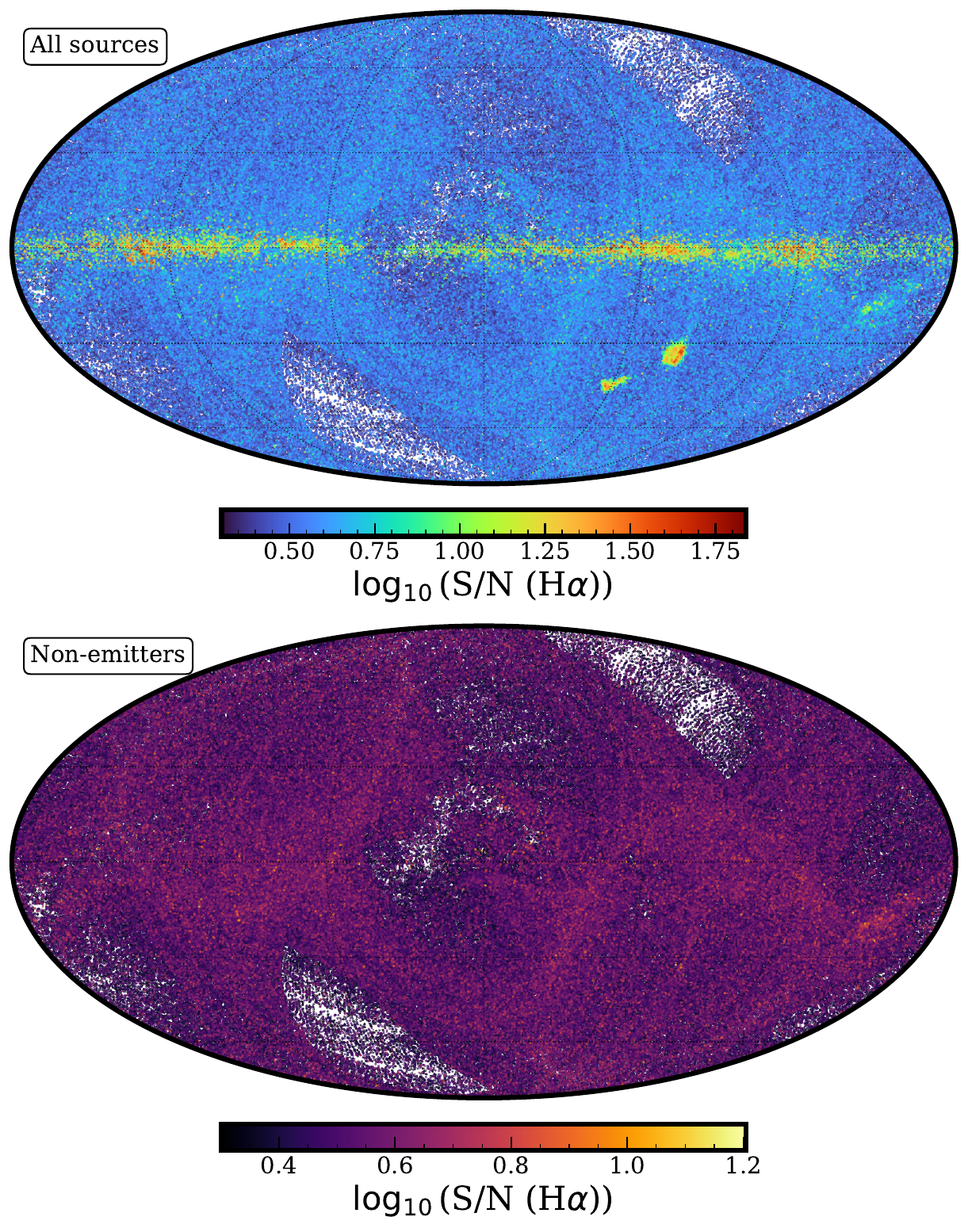}
    \caption{Signal-to-noise ratio of the \Ha equivalent width as a function sky position. \textit{Top}: On-sky distribution of the S/N (\Ha) for all 556\,100 sources in the initial sample. \textit{Bottom}: Same but only for false positives.
    }
    \label{fig:scanning_law_snr}
\end{figure}

\begin{figure}
    \centering
\includegraphics[width=0.959\columnwidth, height=10cm]{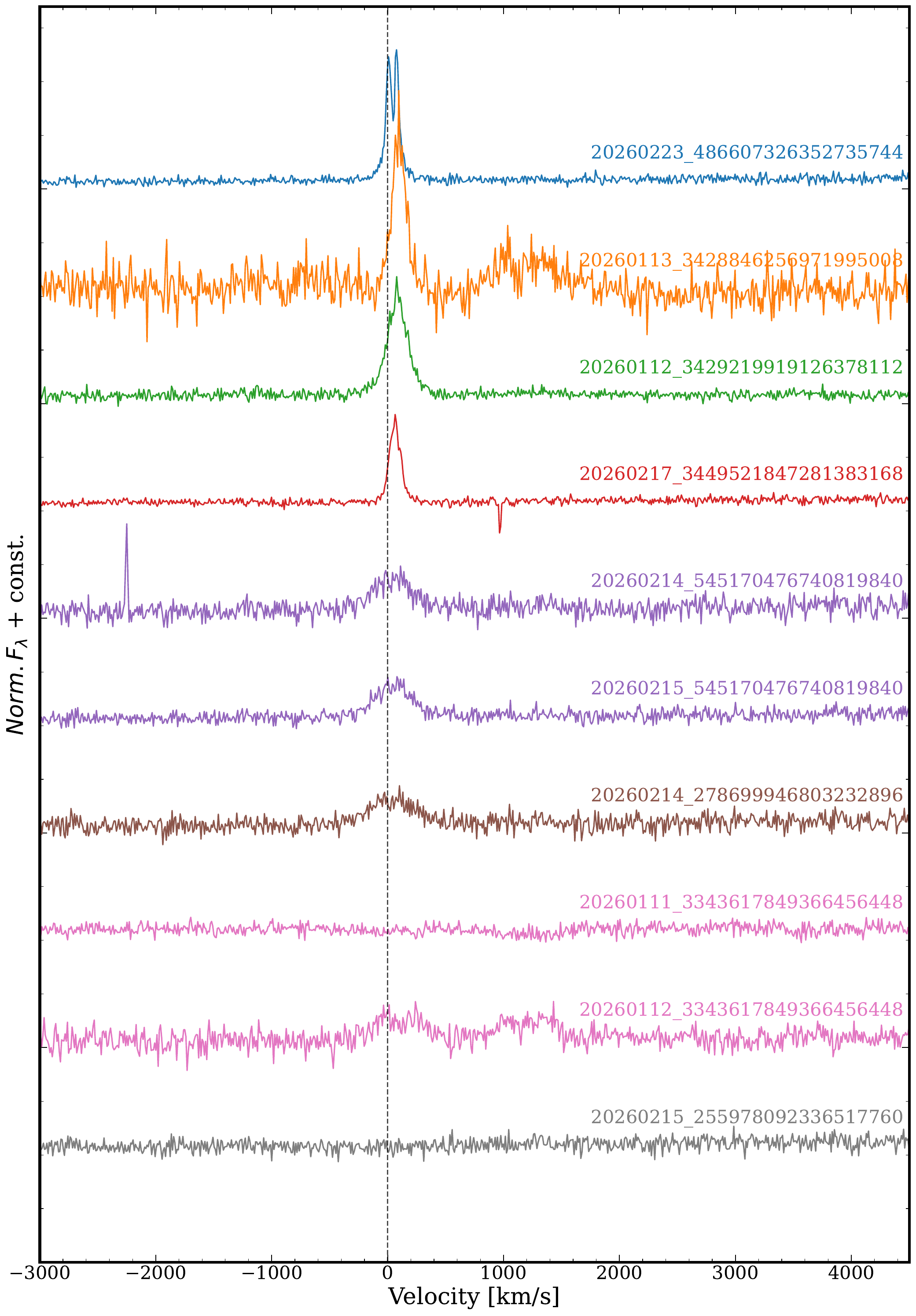}
    \caption{Same as Fig.~\ref{fig:tjo_halpha_emitters} but for lower S/N of the observed TJO spectra.
    }
    \label{fig:tjo_halpha_emitters_low_SNR}
\end{figure}

\begin{table*}[t]
\caption{Comparison of our catalogue with literature catalogues.}
\centering
\begin{tabular}{lccccc}
\hline
\hline
Reference & Stellar Population & \shortstack{Sources in Ref.\\{\small (with XP Spectra)}} & Total Matches & manual\_spectral\_label & Matches \\
\hline
\noalign{\vskip 4pt}
\citet{Vioque2020} & Be & 693 (652) & 449 (68.87\%) & {\tt Be?} & 445 \\
 &  &  &  & {\tt WR} & 3 \\
 &  &  &  & {\tt PN} & 1 \\
\rule{0pt}{1.05ex} & \rule{0pt}{1.05ex} & \rule{0pt}{1.05ex} & \rule{0pt}{1.05ex} & \rule{0pt}{1.05ex} & \rule{0pt}{1.05ex} \\
\citet{Garcia-Moreno2026} & Be & 67 (67) & 6 (8.96\%) & {\tt Be?} & 6 \\
\rule{0pt}{1.05ex} & \rule{0pt}{1.05ex} & \rule{0pt}{1.05ex} & \rule{0pt}{1.05ex} & \rule{0pt}{1.05ex} & \rule{0pt}{1.05ex} \\
\citet{Chojnowski2017} & Be & 213 (191) & 50 (26.18\%) & {\tt Be?} & 50 \\
\rule{0pt}{1.05ex} & \rule{0pt}{1.05ex} & \rule{0pt}{1.05ex} & \rule{0pt}{1.05ex} & \rule{0pt}{1.05ex} & \rule{0pt}{1.05ex} \\
\citet{Tan2025} & Be & 2503 (2038) & 47 (2.31\%) & {\tt Be?} & 47 \\
\rule{0pt}{1.05ex} & \rule{0pt}{1.05ex} & \rule{0pt}{1.05ex} & \rule{0pt}{1.05ex} & \rule{0pt}{1.05ex} & \rule{0pt}{1.05ex} \\
\citet{Shenar2024} & Massive Stars in SMC & 929 (925) & 69 (7.46\%) & {\tt Be?} & 68 \\
 &  &  &  & {\tt NA} & 1 \\
\rule{0pt}{1.05ex} & \rule{0pt}{1.05ex} & \rule{0pt}{1.05ex} & \rule{0pt}{1.05ex} & \rule{0pt}{1.05ex} & \rule{0pt}{1.05ex} \\
\citet{Neiner2011} & Be & 2451 (2215) & 1132 (51.11\%) & {\tt Be?} & 1117 \\
 &  &  &  & {\tt NA} & 10 \\
 &  &  &  & {\tt Herbig?} & 3 \\
 &  &  &  & {\tt Mix} & 1 \\
\rule{0pt}{1.05ex} & \rule{0pt}{1.05ex} & \rule{0pt}{1.05ex} & \rule{0pt}{1.05ex} & \rule{0pt}{1.05ex} & \rule{0pt}{1.05ex} \\
\citet{An2026} & Be & 504 (470) & 207 (44.04\%) & {\tt Be?} & 207 \\
\rule{0pt}{1.05ex} & \rule{0pt}{1.05ex} & \rule{0pt}{1.05ex} & \rule{0pt}{1.05ex} & \rule{0pt}{1.05ex} & \rule{0pt}{1.05ex} \\
\citet{Shridharan2021} & H$\alpha$ Emitters & 3017 (2759) & 426 (15.44\%) & {\tt Be?} & 422 \\
 &  &  &  & {\tt FGK?} & 2 \\
 &  &  &  & {\tt M} & 1 \\
 &  &  &  & {\tt CV?MB} & 1 \\
\rule{0pt}{1.05ex} & \rule{0pt}{1.05ex} & \rule{0pt}{1.05ex} & \rule{0pt}{1.05ex} & \rule{0pt}{1.05ex} & \rule{0pt}{1.05ex} \\
\citet{Tan_2025} & H$\alpha$ Emitters & 23119 (16846) & 689 (4.09\%) & {\tt Be?} & 646 \\
 &  &  &  & {\tt M} & 31 \\
 &  &  &  & {\tt Herbig?} & 8 \\
 &  &  &  & {\tt NA} & 2 \\
 &  &  &  & {\tt [WR]} & 1 \\
 &  &  &  & {\tt FGK?} & 1 \\
\rule{0pt}{1.05ex} & \rule{0pt}{1.05ex} & \rule{0pt}{1.05ex} & \rule{0pt}{1.05ex} & \rule{0pt}{1.05ex} & \rule{0pt}{1.05ex} \\
\citet{Yu2025} & H$\alpha$ Emitters & 45220 (33004) & 462 (1.4\%) & {\tt Be?} & 438 \\
 &  &  &  & {\tt M} & 17 \\
 &  &  &  & {\tt Herbig?} & 6 \\
 &  &  &  & {\tt [WR]} & 1 \\
\rule{0pt}{1.05ex} & \rule{0pt}{1.05ex} & \rule{0pt}{1.05ex} & \rule{0pt}{1.05ex} & \rule{0pt}{1.05ex} & \rule{0pt}{1.05ex} \\
\citet{Merc2019, Merc2026} & Symbiotic & 1254 (956) & 322 (33.68\%) & {\tt M} & 152 \\
 &  &  &  & {\tt NA} & 97 \\
 &  &  &  & {\tt Be?} & 51 \\
 &  &  &  & {\tt Herbig?} & 9 \\
 &  &  &  & {\tt C?} & 8 \\
 &  &  &  & {\tt CV?MB} & 3 \\
 &  &  &  & {\tt PN} & 1 \\
\rule{0pt}{1.05ex} & \rule{0pt}{1.05ex} & \rule{0pt}{1.05ex} & \rule{0pt}{1.05ex} & \rule{0pt}{1.05ex} & \rule{0pt}{1.05ex} \\
\citet{Shi2026} & QSO & 123342 (123342) & 707 (0.57\%) & {\tt Be?} & 218 \\
 &  &  &  & {\tt QSO} & 204 \\
 &  &  &  & {\tt Mix} & 172 \\
 &  &  &  & {\tt NA} & 29 \\
 &  &  &  & {\tt Herbig?} & 11 \\
 &  &  &  & {\tt FGK?} & 11 \\
 &  &  &  & {\tt WR} & 10 \\
 &  &  &  & {\tt [WR]} & 1 \\
 &  &  &  & {\tt M} & 1 \\
\rule{0pt}{1.05ex} & \rule{0pt}{1.05ex} & \rule{0pt}{1.05ex} & \rule{0pt}{1.05ex} & \rule{0pt}{1.05ex} & \rule{0pt}{1.05ex} \\
\hline
\end{tabular}
\tablefoot{\referee{We note that the total matches with our catalogue can be more than the sum of matches with individual spectral classes due to some sources with {\tt classification\_flag} = 0 in our catalogue, and thus not being counted in the individual classes.}}
\label{tab:lit_comparison}
\end{table*}

\begin{table*}[t]
\ContinuedFloat
\caption{Continued.}
\centering
\begin{tabular}{lccccc}
\hline
\hline
Reference & Stellar Population & \shortstack{Sources in Ref.\\{\small (with XP Spectra)}} & Total Matches & manual\_spectral\_label & Matches \\
\hline
\noalign{\vskip 4pt}
\citet{Hou2020} & CV & 284 (170) & 32 (18.82\%) & {\tt Be?} & 17 \\
 &  &  &  & {\tt Mix} & 10 \\
 &  &  &  & {\tt M} & 2 \\
 &  &  &  & {\tt CV?MB} & 2 \\
 &  &  &  & {\tt FGK?} & 1 \\
\rule{0pt}{1.05ex} & \rule{0pt}{1.05ex} & \rule{0pt}{1.05ex} & \rule{0pt}{1.05ex} & \rule{0pt}{1.05ex} & \rule{0pt}{1.05ex} \\
\citet{Rodriguez2025} & CV & 590 (310) & 93 (30.0\%) & {\tt Be?} & 42 \\
 &  &  &  & {\tt Mix} & 41 \\
 &  &  &  & {\tt CV?MB} & 8 \\
 &  &  &  & {\tt NA} & 1 \\
\rule{0pt}{1.05ex} & \rule{0pt}{1.05ex} & \rule{0pt}{1.05ex} & \rule{0pt}{1.05ex} & \rule{0pt}{1.05ex} & \rule{0pt}{1.05ex} \\
\citet{Scholch2026} & \shortstack{Young Stellar Bridge\\(LMC--SMC)} & 12864 (4637) & 440 (9.49\%) & {\tt Be?} & 440 \\
\rule{0pt}{1.05ex} & \rule{0pt}{1.05ex} & \rule{0pt}{1.05ex} & \rule{0pt}{1.05ex} & \rule{0pt}{1.05ex} & \rule{0pt}{1.05ex} \\
\citet{Vioque2020} & Pre-MS & 8470 (4576) & 1126 (24.61\%) & {\tt Be?} & 619 \\
 &  &  &  & {\tt M} & 221 \\
 &  &  &  & {\tt Herbig?} & 214 \\
 &  &  &  & {\tt NA} & 51 \\
 &  &  &  & {\tt FGK?} & 5 \\
 &  &  &  & {\tt CV?MB} & 3 \\
 &  &  &  & {\tt PN} & 1 \\
 &  &  &  & {\tt WR} & 1 \\
\rule{0pt}{1.05ex} & \rule{0pt}{1.05ex} & \rule{0pt}{1.05ex} & \rule{0pt}{1.05ex} & \rule{0pt}{1.05ex} & \rule{0pt}{1.05ex} \\
\citet{Roulston2025} & C & 43574 (43574) & 20 (0.05\%) & {\tt C?} & 16 \\
 &  &  &  & {\tt Be?} & 2 \\
 &  &  &  & {\tt NA} & 1 \\
 &  &  &  & {\tt FGK?} & 1 \\
\rule{0pt}{1.05ex} & \rule{0pt}{1.05ex} & \rule{0pt}{1.05ex} & \rule{0pt}{1.05ex} & \rule{0pt}{1.05ex} & \rule{0pt}{1.05ex} \\
\citet{Rosslowe2015} & WR & 705 (348) & 156 (44.83\%) & {\tt WR} & 101 \\
 &  &  &  & {\tt Be?} & 50 \\
 &  &  &  & {\tt NA} & 3 \\
 &  &  &  & {\tt Herbig?} & 1 \\
 &  &  &  & {\tt M} & 1 \\
\rule{0pt}{1.05ex} & \rule{0pt}{1.05ex} & \rule{0pt}{1.05ex} & \rule{0pt}{1.05ex} & \rule{0pt}{1.05ex} & \rule{0pt}{1.05ex} \\
\citet{Acker2003} & [WR] & 121 (64) & 15 (23.44\%) & {\tt [WR]} & 7 \\
 &  &  &  & {\tt WR} & 4 \\
 &  &  &  & {\tt Be?} & 3 \\
 &  &  &  & {\tt NA} & 1 \\
\rule{0pt}{1.05ex} & \rule{0pt}{1.05ex} & \rule{0pt}{1.05ex} & \rule{0pt}{1.05ex} & \rule{0pt}{1.05ex} & \rule{0pt}{1.05ex} \\
\citet{Hodgkin2021} & Gaia\_Alerts & 7561 (938) & 74 (7.89\%) & {\tt Be?} & 30 \\
 &  &  &  & {\tt Mix} & 23 \\
 &  &  &  & {\tt M} & 10 \\
 &  &  &  & {\tt CV?MB} & 4 \\
 &  &  &  & {\tt Herbig?} & 3 \\
 &  &  &  & {\tt NA} & 3 \\
 &  &  &  & {\tt FGK?} & 1 \\
\rule{0pt}{1.05ex} & \rule{0pt}{1.05ex} & \rule{0pt}{1.05ex} & \rule{0pt}{1.05ex} & \rule{0pt}{1.05ex} & \rule{0pt}{1.05ex} \\
\citet{Delfini2025} & YSO & 1945 (1241) & 723 (58.26\%) & {\tt M} & 563 \\
 &  &  &  & {\tt Be?} & 75 \\
 &  &  &  & {\tt Herbig?} & 41 \\
 &  &  &  & {\tt NA} & 27 \\
 &  &  &  & {\tt CV?MB} & 10 \\
\hline
\end{tabular}
\end{table*}

Figure~\ref{fig:confusion_matrix_manual_spectral_labels_vs_gaiadr3_esp_hs_labels} shows the direct comparison of our spectral labels with the classes reported by the \Gaia DR3 ESP-HS (Extended Stellar Parametrizer for Hot Stars) pipeline \citep{Fouesneau2023}. Figure~\ref{fig:scanning_law_snr} visualises the sky distribution of the signal-to-noise ratios of the \Ha equivalent width, S/N (\Ha). The top panel demonstrates that this distribution is mainly driven by the distribution of various stellar populations in the Galaxy (e.g. {\tt Be?} stars are mostly confined to the Galactic Plane), while the bottom panel highlights the imprint of the \Gaia DR3 scanning law on the distribution of spurious sources. Figure~\ref{fig:tjo_halpha_emitters_low_SNR} shows the TJO/ARES spectra of the low-S/N sample of the Be star candidates followed up in Sect.~\ref{subsec:montsec}, in the same fashion as Fig.~\ref{fig:tjo_halpha_emitters}. Finally, Tab.~\ref{tab:lit_comparison} summarises our comparisons with the literature catalogues of \Ha emitters discussed in Sect.~\ref{sec:literature}.

\end{appendix}

\end{document}